\documentclass{WileYMSP-template}
\usepackage[T1]{fontenc}

\usepackage{accents}
\usepackage{hyperref}
\usepackage{amssymb}
\usepackage{graphicx} 
\usepackage{caption}
\usepackage{subfig}

\begin{document}

\pagestyle{fancy}
\rhead{\includegraphics[width=2.5cm]{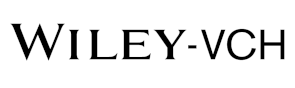}}

\title{Constant-roll inflation in the generalized SU(2) Proca theory}

\maketitle


\author{Juan C. Garnica}
\author{L. Gabriel G\'omez}
\author{Andr\'es A. Navarro}
\author{Yeinzon Rodr\'iguez*}


\dedication{Dedicated to Carlos J. Quimbay Herrera, Marta A. Losada Falk, and David H. Lyth}

\reportnumber{PI/UAN-2021-699FT}

\begin{affiliations}
Juan C. Garnica\\
Escuela  de  F\'isica,  Universidad  Industrial  de  Santander, \\ 
Ciudad  Universitaria,  Bucaramanga  680002,  Colombia\\
Email Address: juancagarnica@hotmail.com\\

Dr. L. Gabriel G\'omez\\
Departamento  de  F\'isica,  Universidad  de Santiago de Chile, \\ 
Avenida V\'{\i}ctor Jara 3493, Estaci\'on Central, 9170124, Santiago, Chile\\
Email Address: gabriel.gomez.d@usach.cl\\

Dr. Andr\'es A. Navarro\\
Departamento de Ciencias B\'asicas,  Universidad Santo Tom\'as, \\ 
Cra 17 \# 9-27,  Bucaramanga  680002,  Colombia\\
Email Address: andres.navarro@ustabuca.edu.co\\

Prof. Yeinzon Rodr\'iguez\\
Centro de Investigaciones en Ciencias B\'asicas y Aplicadas, Universidad Antonio Nari\~no, \\ 
Cra 3 Este \# 47A-15, Bogot\'a D.C. 110231, Colombia\\
Escuela  de  F\'isica,  Universidad  Industrial  de  Santander, \\ 
Ciudad  Universitaria,  Bucaramanga  680002,  Colombia\\
Email Address: yeinzon.rodriguez@uan.edu.co

\end{affiliations}



\begin{abstract}

The generalized SU(2) Proca theory (GSU2P) is a variant of the well known generalized Proca theory where the vector field belongs to the Lie algebra of the SU(2) group of global transformations under which the action is made invariant. New interesting possibilities arise in this framework because of the existence of new interactions of purely non-Abelian character and new configurations of the vector field resulting in spatial spherical symmetry and the cosmological dynamics being driven by the propagating degrees of freedom. We study the two-dimensional phase space of the system that results when the cosmic triad configuration is employed in the Friedmann-Lemaitre-Robertson-Walker background and find an attractor curve whose attraction basin both covers almost all the allowed region and does not include a Big-Bang singularity. Such an attractor curve corresponds to a primordial inflationary solution that has the following characteristic properties: 1.) it is a de Sitter solution whose Hubble parameter is regulated by a generalized version of the SU(2) group coupling constant, 2.) it is constant-roll including, as a limiting case, the slow-roll variety, 3.) a number of e-folds $N > 60$ is easily reached, 4.) it has a graceful exit into a radiation dominated period powered by the canonical kinetic term of the vector field and the Einstein-Hilbert term. The free parameters of the action are chosen such that the tensor sector of the theory is the same as that of general relativity at least up to second-order perturbations, thereby avoiding the presence of ghost and Laplacian instabilities in the tensor sector as well as making the gravity waves propagate at light speed. This is a proof of concept of the interesting properties we could find in this scenario when the coupling constants be replaced by general coupling functions and more terms be discovered in the GSU2P.

\end{abstract}


\section{Introduction}
Unless the evidence \cite{Reuter:2001ag,Litim:2003vp,Falls:2014tra,Eichhorn:2018yfc} that General Relativity (GR) is non-perturbatively renormalizable becomes decisive, Einstein gravity continues to be an effective theory \cite{Donoghue:1994dn,Burgess:2003jk}.  The search for a modified gravity theory (see Refs. \cite{Nojiri:2006ri,Sotiriou:2008rp,DeFelice:2010aj,Capozziello:2011et,Clifton:2011jh,Hinterbichler:2011tt,deRham:2014zqa,Schmidt-May:2015vnx,Heisenberg:2018vsk} for some reviews), effective as well, is then completely justified despite the experimental success of GR.  There exist different ways to approach the problem like adding higher-order curvature invariants to the action, considering also geometrical structures other than the pseudo-Riemmanian curvature, or incrementing the gravitational degrees of freedom.  It is well known that this path is full of obstacles making particularly difficult to reconcile renormalizability with unitarity in the search for a fundamental theory.  The unitarity is intimately related to the absence of ghosts and instabilities \cite{Woodard:2006nt,Woodard:2015zca} and, therefore, to the existence of constraints that guarantee the propagation of the right number of degrees of freedom.  One procedure general enough to produce these constraints, at least those primary, is to degenerate the kinetic Lagrangian.  This procedure has been applied when the additional gravitational degree of freedom is of scalar nature, giving existence to the so called Degenerate Higher-Order Scalar-Tensor theory (DHOST) \cite{Langlois:2015cwa,Achour:2016rkg,BenAchour:2016fzp,Crisostomi:2016czh}, and also when such a degree of freedom is of vector nature, giving existence to the so called Extended Vector-Tensor theory (EVT) \cite{Kimura:2016rzw}.  Another procedure, less general but with a longer history, is to construct the action so that the field equations are second order, thereby avoiding the Ostrogradski instability \cite{Ostrogradsky:1850fid}.  It was Horndeski in the 70's who gave us what is understandably called the Horndeski theory \cite{Horndeski:1974wa} (nowadays also called the (generalized) Galileon theory \cite{Nicolis:2008in,Deffayet:2009mn,Deffayet:2009wt,Deffayet:2011gz,Kobayashi:2011nu,Deffayet:2013lga,Rodriguez:2017ckc,Kobayashi:2019hrl}) and that describes gravity composed of a purely geometrical degree of freedom (the curvature) together with a scalar field whose field equations are second order.  Horndeski himself gave us the analogous version when the scalar field is replaced by a vector field enjoying a U(1) internal gauge symmetry, having in mind the electromagnetism \cite{Horndeski:1976gi}.  

A lot of effort has been put in the exploration of the fundamental, cosmological, and astrophysical consequences of the Horndeski theory (see e.g. Refs. \cite{Deffayet:2013lga,Kobayashi:2019hrl}) but it had to face a very severe obstacle:  the prediction of anomalous speed for the gravity waves \cite{Bettoni:2016mij} in huge disagreement with the observations \cite{Baker:2017hug,Creminelli:2017sry,Sakstein:2017xjx,Ezquiaga:2017ekz,Wang:2017rpx,TheLIGOScientific:2017qsa,GBM:2017lvd,LIGOScientific:2017zic}.  This obstacle is saved in the framework of the DHOST theory \cite{Langlois:2017dyl} but, in our opinion, it has been enough for scientists to have lost much of the interest in modified gravity theories.  A lot of effort has been also put in the study of the vector-tensor version of the Horndeski theory, what is called the generalized Proca theory (GP).  This theory is, actually, the generalization of the vector-tensor theory given to us by Horndeski \cite{Horndeski:1976gi} to the case when there is no internal gauge symmetry.  Having been developed by Tasinato \cite{Tasinato:2014eka} and Heisenberg \cite{Heisenberg:2014rta,Jimenez:2016isa}, it was later completed by Allys, Peter, and Rodr\'{\i}guez \cite{Allys:2015sht,Allys:2016jaq} (see also Ref. \cite{Rodriguez:2017ckc}) who discovered an extra Lagrangian piece with trivial longitudinal-mode interactions.  Interesting results were found, both in the astrophysical and cosmological avenues, starting from the discovery of a self-accelerating cosmological solution powered by the non-propagating mode \cite{Tasinato:2014eka,DeFelice:2016yws,Geng:2021jso}.  It is quite unexpected that the non-propagating degree of freedom can give way to some dynamics but, of course, it is very interesting and, anyway, the only possibility since the propagating modes must be turned off if we want a perfectly homogeneous and isotropic universe at the background level.  The GP stood as a very serious alternative until the observations about the gravity waves speed teared down most of its sectors \cite{Baker:2017hug} (curiously, the extra Lagrangian piece described above predicts the right speed).  The EVT, analogously to its scalar version, the DHOST, is able to save the day by providing extra terms that give way to the correct gravity waves speed.

But are the GP and the EVT the last words in the framework of the vector-tensor theories?  Anybody might think that adding a global SU(2) internal symmetry to the GP leads to a very similar theory, with essentially the same implications, although, technically, being much more complicated.  There are, however, at least three reasons why this is not the right reasoning:
\begin{enumerate}
\item There exist terms in the action of the SU(2) version of purely non-Abelian nature, i.e. terms for which there do not exist their counterparts in the GP.  Likewise, there exist terms in the action of the GP that are incompatible with their generalization to the SU(2) version.
\item Not only the non-propagating degree of freedom but also the propagating ones can be turned on without spoiling the isotropy.  In fact, there exist four different field configurations, three of them where the non-propagating degree of freedom is turned off, and their linear combinations that enjoy spatial spherical symmetry \cite{Witten:1976ck,Sivers:1986kq,Forgacs:1979zs}.  One of these linear combinations is the well known cosmic triad \cite{ArmendarizPicon:2004pm}.
\item From the gauge fundamental interactions in particle physics, we know well that the non-Abelian interactions exhibit a rich phenomenology which is very characteristic and distinctive from the Abelian interactions'.  For instance, gauge mediators in the any of the nuclear forces interact with each other whereas photons never do it. 
\end{enumerate}
These reasons led to the formulation of what is called the generalized SU(2) Proca theory (GSU2P) \cite{Allys:2016kbq} (see also Ref. \cite{Jimenez:2016upj} and for a short review see Ref. \cite{Rodriguez:2017ckc}) which was later on reconstructed \cite{GallegoCadavid:2020dho} to take into account not only the primary constraints but also the secondary constraints that close the constraint algebra \cite{ErrastiDiez:2019ttn,ErrastiDiez:2019trb} (see also Ref. \cite{ErrastiDiez:2020dux}).  Such reconstruction paid careful attention to the covariantization process so that the beyond SU(2) Proca terms (parts of what would be an extended version of the GSU2P) were unveiled following the technique devised in Ref \cite{GallegoCadavid:2019zke}.

It is the purpose of this paper to investigate the cosmological implications of the GSU2P, particularly on the primordial inflationary period.  Some preliminar research on the impact of this theory on the late inflationary period was carried out in Ref. \cite{Rodriguez:2017wkg} and the stability conditions to reach this purpose in Ref. \cite{Gomez:2019tbj}.  Here we will concentrate on the background evolution during primordial inflation, identifying the inflaton with the single degree of freedom that results when the cosmic triad configuration is employed.  Our action will be Einstein-Hilbert plus Yang-Mills (Y-M) plus the parity-conserving $\mathcal{L}_{4,2}$ Lagrangian pieces of the GSU2P plus those $\mathcal{L}_2$ pieces that are composed of either two derivatives and two vector fields or four vector fields.  The $\mathcal{L}_{4,2}$ Lagrangian pieces of the GSU2P give way to an almost universal attractor curve in phase space along which the slow-roll parameter $\epsilon$ vanishes.  Thus, for most of the initial conditions, the system ends up experiencing a quasi-de Sitter inflationary period which easily lasts more than 60 e-folds.  Such a period has the property of being constant-roll \cite{Motohashi:2014ppa,Motohashi:2017vdc,Motohashi:2019tyj} because $\ddot{\psi}/H \dot{\psi}$ (as well as the $\eta$ slow-roll parameter), where $\psi$ is the single degree of freedom that describes the cosmic triad and $H$ is the Hubble parameter, is constant along the attractor curve;  slow-roll inflation appears as a limiting case in this scenario.  Another property is that the energy scale during inflation, given by $H$, may take any value depending on a generalized version of the SU(2) group coupling constant.  On the other hand, the Y-M term has the role of producing a graceful exit of inflation into a radiation-dominated period when $\psi$ becomes small enough.  Finally, the Einstein-Hilbert term guarantees as well a reasonable cosmological evolution during the radiation-dominated period and the recovery of the Einstein gravity once the vector field has decayed.

An appreciable quantity of free parameters can make quite intractable the analysis of this scenario.  In order to get a clear picture of what could happen in the most general situation, we have notoriously reduced the number of free parameters by analyzing the perturbative tensor sector of the theory and imposing conditions so that it takes the form shown in GR, at least up to second order.  Thus, the ghost and Laplacian instabilities problems in the tensor sector are solved while making the gravity waves propagate at light speed.  Of course, this is more than what is required in a perturbative analysis but perfectly fits our intention of easying the work and uncovering some interesting aspects of the dynamics that are likely present in the most general situation.  At such, this paper becomes a proof of concept of what we expect to find when the GSU2P be enlarged with new terms and the coupling constants be replaced by general coupling functions (the GSU2P \cite{Allys:2016kbq,GallegoCadavid:2020dho} was partially constructed because of the difficulty at handling hundreds of Lorentz-invariant and SU(2) group-invariant Lagrangian building blocks).

The layout of this paper is the following.  After presenting the action to study and the field profile in Section \ref{model}, a dynamical system is constructed in Section \ref{dynsys} and the respective field equations are formulated in terms of the dynamical system variables.  Thus, the existence of an attractor curve (actually a straight line in the two-dimensional phase space) corresponding to a de Sitter inflationary period is revealed in Section \ref{inflation}.  To analyze the properties of this inflationary period and the other features of the scenario under study, it is easier to perform a reduction of the set of free parameters, described in Section \ref{reduction}, by analyzing the tensor sector of the theory and making it behave as in GR at least up to second order;  the side (positive) effect of this is that the scenario becomes free of ghosts and Laplacian instabilities.  The graceful exit of inflation into a radiation dominated radiation and its impact on the determination of the available parameter space are studied in Sections \ref{gracexit} and \ref{avalparspace} whereas the constant-roll characteristic of the inflationary period, its limiting cases, and the past singularities are discussed in Section \ref{crolls}.  The comparison of this scenario with others available in the literature is carried out in Section \ref{comparison}.  Finally, the conclusions are presented in Section \ref{conclusions}.  Throughout the text, Greek indices are space-time indices and run from 0 to 3, while the Latin indices $a,b,c$ are internal SU(2) group indices and run from 1 to 3.  The sign convention is the (+++) according to Misner, Thorne, and Wheeler \cite{Misner:1974qy}.

\section{The GSU2P and the scenario to study} \label{model}

The GSU2P is a generalization of the GP to the case where the vector field belongs to the Lie algebra of the SU(2) group.  The action is then globally invariant under this group of transformations.  As in the GP \cite{Deffayet:2013tca}, gauge invariance is abandoned in order to have more in the action than just those terms that are built with gauge-invariant contractions of the gauge-field strength tensor with itself (e.g. the Maxwell or Y-M kinetic terms). The theory was constructed in Ref. \cite{Allys:2016kbq} (see also Ref. \cite{Jimenez:2016upj}) implementing the primary constraint-enforcing relation only and removing several terms which were redundant in flat spacetime.  After the works in Refs. \cite{ErrastiDiez:2019ttn,ErrastiDiez:2019trb}, it became clear that a non-trivial secondary constraint-enforcing relation existed and that ought to be implemented.  This was, however, almost impossible to carry out from the results in Ref. \cite{Allys:2016kbq} since several of the removed redundant terms did not satisfy this secondary constraint-enforcing relation, forcing the authors of Ref. \cite{GallegoCadavid:2020dho} to reconstruct the theory from scratch.  Anyway, removing these terms before covariantizing implied the lost of information since some of them are not redundant anymore in curved spacetime and, indeed, are intimately related to what are called the beyond Proca SU(2) terms \cite{GallegoCadavid:2019zke}.  The theory was reconstructed in Ref. \cite{GallegoCadavid:2020dho} finding in turn the beyond Proca SU(2) terms.  The action of the theory is the following:
\begin{equation}
S = \int d^4 x \ \sqrt{-g} \ \left(\mathcal{L}_{E-H} + \mathcal{L}_2 + \alpha_{4,0} \mathcal{L}_{4,0} + \mathcal{L}_{4,2} + \frac{\alpha_{5,0}}{m_P^2} \tilde{\mathcal{L}}_{5,0} \right) \,, \label{GSU2Paction}
\end{equation}
where
\begin{eqnarray}
\mathcal{L}_{E-H} &\equiv& \frac{m_P^2}{2} R \,, \\
\mathcal{L}_2 &\equiv& \mathcal{L}_2(A_{\mu \nu}^a,B_\mu^a) \,, \\
\mathcal{L}_{4,0} &\equiv& G_{\mu \nu} B^{\mu a} B^\nu_a \,, \label{L20} \\
\mathcal{L}_{4,2} &\equiv& \sum_{i = 1}^6 \frac{\alpha_i}{m_P^2} \mathcal{L}_{4,2}^i + \sum_{i = 1}^4 \frac{\tilde{\alpha}_i}{m_P^2} \tilde{\mathcal{L}}_{4,2}^i \,, \label{L42} \\
\tilde{\mathcal{L}}_{5,0} &\equiv& B^{\nu a} R^\sigma_{\ \ \nu \rho \mu} B_\sigma^b \tilde{A}^{\mu \rho c} \epsilon_{abc} \,, \label{L50}
\end{eqnarray}
and
\begin{eqnarray}
\mathcal{L}_{4,2}^1 &\equiv& (B_b \cdot B^b) [S^{\mu a}_\mu S^\nu_{\nu a} - S^{\mu a}_\nu S^\nu_{\mu a}] + 2 (B_a \cdot B_b) [S^{\mu a}_\mu S^{\nu b}_\nu - S^{\mu a}_\nu S^{\nu b}_\mu] \,, \nonumber \\
\mathcal{L}_{4,2}^2 &\equiv& A_{\mu \nu}^a S^{\mu b}_\sigma B^\nu_a B^\sigma_b - A_{\mu \nu}^a S^{\mu b}_\sigma B^\nu_b B^\sigma_a + A_{\mu \nu}^a S^{\rho b}_\rho B^\mu_a B^\nu_b \,, \nonumber \\
\mathcal{L}_{4,2}^3 &\equiv& B^{\mu a} R^\alpha_{\ \ \sigma \rho \mu} B_{\alpha a} B^{\rho b}  B^\sigma_b + \frac{3}{4} (B_b \cdot B^b) (B^a \cdot B_a) R \,, \nonumber \\
\mathcal{L}_{4,2}^4 &\equiv& [(B_b \cdot B^b) (B^a \cdot B_a) + 2 (B_a \cdot B_b) (B^a \cdot B^b)] R \,, \nonumber \\
\mathcal{L}_{4,2}^5 &\equiv& G_{\mu \nu} B^{\mu a} B^\nu_a (B^b \cdot B_b) \,, \nonumber \\
\mathcal{L}_{4,2}^6 &\equiv& G_{\mu \nu} B^{\mu a} B^{\nu b} (B_a \cdot B_b) \,, \label{L42LPa}
\end{eqnarray}
\begin{eqnarray}
\tilde{\mathcal{L}}_{4,2}^1 &\equiv& -2 A_{\mu \nu}^a S^{\mu b}_\sigma B_{\alpha a} B_{\beta b} \epsilon^{\nu \sigma \alpha \beta} + S^a_{\mu \nu} S^{\nu b}_\sigma B_{\alpha a} B_{\beta b} \epsilon^{\mu \sigma \alpha \beta} \,, \nonumber \\
\tilde{\mathcal{L}}_{4,2}^2 &\equiv& A_{\mu \nu}^a S^{\mu b}_\sigma B_{\alpha a} B_{\beta b} \epsilon^{\nu \sigma \alpha \beta} - \tilde{A}^{\alpha \beta}_a S^b_{\rho \alpha} B^{\rho a} B_{\beta b} + \tilde{A}^{\alpha \beta}_a S^\rho_{\rho b} B_\alpha^a B_\beta^b \,, \nonumber \\
\tilde{\mathcal{L}}_{4,2}^3 &\equiv& B_\beta^b R^\alpha_{\ \ \sigma \rho \mu} B_\alpha^a (B_a  \cdot B_b) \epsilon^{\mu \rho \sigma \beta} \,, \nonumber \\
\tilde{\mathcal{L}}_{4,2}^4 &\equiv& B_{\beta a} R^\alpha_{\ \ \sigma \rho \mu} B_\alpha^a (B^b  \cdot B_b) \epsilon^{\mu \rho \sigma \beta} \,. \label{L42LPb}
\end{eqnarray}
In these expressions, $g$ is the determinant of the metric, $m_P$ is the reduced Planck mass, $R$ is the Ricci scalar, $B_\mu^a$ is the vector field that belongs to the Lie algebra of the SU(2) group, $A_{\mu \nu}^a \equiv \nabla_\mu B_\nu^a - \nabla_\nu B_\mu^a$ is the Abelian version of the non-Abelian gauge-field strength tensor $F_{\mu \nu}^a$ where $\nabla_\mu$ is the space-time covariant derivative operator, $S_{\mu \nu}^a \equiv \nabla_\mu B_\nu^a + \nabla_\nu B_\mu^a$ is the symmetric version of $A_{\mu \nu}^a$, $G_{\mu \nu}$ is the Einstein tensor, $R^\sigma_{\ \ \nu \rho \mu}$ is the Riemann tensor, $\tilde{A}^{\mu \nu}_a \equiv \frac{1}{2} \epsilon^{\mu \nu \rho \sigma} A_{\rho \sigma a}$ is the Hodge dual of $A^{\mu \nu}_a$, $\epsilon_{abc}$ is the structure-constant tensor of the SU(2) group, $\alpha_{4,0}$, $\alpha_{5,0}$, $\alpha_i$, and $\tilde{\alpha}_i$ are arbitrary dimensionless constants, and $\epsilon^{\mu \nu \rho \sigma}$ is the Levi-Civita tensor of the space-time manifold.  It is worth noting that those Lagrangians with a tilde on top explicitly violate parity while those without a tilde do not.  The exception is $\mathcal{L}_2$ which is an arbitrary function of $A_{\mu \nu}^a$ and $B_\mu^a$, i.e., it can be any scalar built from contractions of the latter two objects with the metric and Levi-Civita tensors of the space-time and SU(2) group manifolds and, therefore, may include both parity-conserving and parity-violating terms.  The notation here has been changed a bit with respect to that in Ref. \cite{GallegoCadavid:2020dho} in order to make things clearer:  $B$, instead of $A$, stands for the vector field, $A$ stands for the unnormalized antisymmetrization of $\nabla_\mu B_\nu^a$, and $S$ stands for its unnormalized symmetrization.

One of the guiding ideas of the scenario we consider in this paper is that the inflationary period has a graceful exit into a radiation dominated period and that the Einstein gravity is recovered once the vector field has decayed.  Therefore, part of the $\mathcal{L}_2$ we consider here is the canonical Y-M kinetic term:
\begin{equation}
\mathcal{L}_2 \ni \mathcal{L}_{Y-M} \equiv -\frac{1}{4}F_{\mu \nu}^a F^{\mu \nu}_a \,,
\end{equation}
where $F_{\mu \nu}^a \equiv \nabla_\mu B_\nu^a - \nabla_\nu B_\mu^a + \tilde{g} \epsilon^a_{\;\; bc} B_\mu^b B_\nu^c$, $\tilde{g}$ being the SU(2) group coupling constant.  On the other hand, the existence of the inflationary period and its graceful exit itself depend noticeably on the terms constructed from two derivatives and two vector fields and on those constructed from four vector fields, i.e., we have to consider those in $\mathcal{L}_{4,2}$, except for the parity-violating Lagrangian pieces because they are inconsistent with an isotropic universe.  Thereby, it is reasonable to consider as well the following Lagrangian pieces that belong to $\mathcal{L}_2$:
\begin{eqnarray}
\mathcal{L}_2^1 &\equiv& (B^a \cdot B_a) (B^b \cdot B_b) \,, \nonumber \\
\mathcal{L}_2^2 &\equiv& (B^a \cdot B^b) (B_a \cdot B_b) \,, \nonumber \\
\mathcal{L}_2^3 &\equiv& A^{\mu \nu}_a A^{\rho \;\;\; a}_{\;\; \nu} B_\mu^b B_{\rho b} \,, \nonumber \\
\mathcal{L}_2^4 &\equiv& A^{\mu \nu}_a A^{\rho \;\;\; b}_{\;\; \nu} B_{\mu b} B_\rho^a \,, \nonumber \\
\mathcal{L}_2^5 &\equiv& A^{\mu \nu}_a A^{\rho \;\;\; b}_{\;\; \nu} B_\mu^a B_{\rho_b} \,, \nonumber \\
\mathcal{L}_2^6 &\equiv& A^{\mu \nu}_a A_{\mu \nu}^a (B_b \cdot B^b) \,, \nonumber \\
\mathcal{L}_2^7 &\equiv& A^{\mu \nu}_a A_{\mu \nu}^b (B^a \cdot B_b) \,,
\end{eqnarray}
and to exclude $\mathcal{L}_{4,0}$ and $\tilde{\mathcal{L}}_{5,0}$.  Thus, the action corresponding to the scenario that is studied in this paper is given by
\begin{equation}
S = \int d^4 x \ \sqrt{-g} \ \left(\mathcal{L}_{E-H} + \mathcal{L}_{Y-M} + \sum_{i = 1}^2 \chi_i \mathcal{L}_2^i + \sum_{i = 3}^7 \frac{\chi_i}{m_P^2} \mathcal{L}_2^i + \sum_{i = 1}^6 \frac{\alpha_i}{m_P^2} \mathcal{L}_{4,2}^i \right) \,, \label{action}
\end{equation}
where the $\chi_i$ are arbitrary dimensionless constants.  It is very easy to observe that some of the terms in the action of the GSU2P, Eq. (\ref{GSU2Paction}), have purely non-Abelian nature and, therefore, their counterparts in the GP do not exist.  Among such terms, the one that is preserved in the action studied in this paper, Eq. (\ref{action}), is $\mathcal{L}_{4,2}^2$. 

As usual, the universe at large is modeled as a four-dimensional, $C^\infty$, connected, Hausdorff, and paracompact manifold which allows it to be endowed with a pseudo-Riemmanian metric compatible with the spatial isotropy with respect to all points, i.e., the Friedmann-Lemaitre-Robertson-Walker (FLRW) metric.  Following the assumption that the spatial sections of the manifold have vanishing intrinsic curvature, the line element according to the FLRW metric has the following form in a Cartesian coordinate system:
\begin{equation}
ds^2 = -dt^2 + a^2(t) [dx^2 + dy^2 + dz^2] \,, \label{metric}
\end{equation}
where $a(t)$ is the scale factor and $t$ is the cosmic time.

Due to its nature, the spatial part of a vector field breaks isotropy, so the only way it can be reconciled with the FLRW universe is by considering it as a time-like vector field whose spatial components vanish.  This has been the strategy when the cosmology has been studied in the GP scenario (see e.g. Refs. \cite{Tasinato:2014eka,DeFelice:2016yws}).  Thus, together with matter fields, the non-propagating degree of freedom is actually the one that drives the cosmological dynamics.  The GSU2P is richer in this sense since, having more than one vector field in physical space, there exist four different field profiles (as well as linear combinations of them) that are compatible with the FLRW metric, three of them where the spatial components are, in general, different to zero \cite{Witten:1976ck,Sivers:1986kq,Forgacs:1979zs}:
\begin{eqnarray}
B_{0 a} (t) &=& b_0 (t) \hat{\overline{r}}_a \,, \nonumber \\
B_{i a} (t) &=& b_1(t) \hat{r}_i \hat{\overline{r}}_a + b_2(t) [\delta_{ia} - \hat{r}_i \hat{\overline{r}}_a] + b_3(t) \epsilon_{ia}^{\;\;\;\, l} \hat{\overline{r}}_l \,.
\end{eqnarray}
In the above expression $i$ and $l$ run from 1 to 3, $b_0(t)$, $b_1(t)$, $b_2(t)$, and $b_3(t)$ are arbitrary functions of the cosmic time, $\hat{\overline{r}}$ is the unit vector in the isospin space pointing in the direction of $\vec{B}_i$, $\hat{r}$ is the unit vector in physical space pointing in the direction of $\vec{B}_a$, $\delta_{ia}$ is the Kronecker delta, and $\epsilon_{ial}$ is the totally antisymmetric symbol.  A simple field configuration, called the cosmic triad \cite{ArmendarizPicon:2004pm}, is obtained when choosing $b_0(t) = b_3(t) = 0$ and $b_1(t) = b_2(t) \neq 0$.  Thus,
\begin{eqnarray}
B_{0 a} (t) &=& 0 \,, \nonumber \\
B_{i a} (t) &=& a(t) \psi(t) \delta_{ia} \,, \label{profile}
\end{eqnarray}
where $\psi(t)$ corresponds to the norm of the physical three-dimensional vector fields.  It is then easy to understand why the cosmic triad configuration leads to a homogeneous and isotropic universe although the vector fields involved are spacelike:  isotropy requires that a rotation in real space has to be locked into a rotation in the isospin space so that the cosmic triad remains invariant.  Moreover, in contrast to the GP scenario, the propagating degrees of freedom in the GSU2P, together with matter fields, are responsible for the cosmological dynamics.  This is the configuration that is employed in the development of this paper.

\section{The dynamical system} \label{dynsys}

Varying the action in Eq. \ref{action} with respect to both the metric and the vector field leads to the field equations.  Their length is enough to fill some pages so they will not be presented here. However, replacing the FLRW metric in Eq. \ref{metric} and the field configuration in Eq. \ref{profile} in the field equations, and defining the new dimensionless variables:
\begin{equation}
x \equiv \frac{\dot{\psi}}{\sqrt{2} m_P H} \,, \hspace{5mm} y \equiv \frac{\psi}{\sqrt{2} m_P} \,, \hspace{5mm} z \equiv \sqrt{\frac{\hat{g}}{2 m_P H}} \psi \,, \label{defdynsys}
\end{equation}
where
\begin{equation}
\hat{g} \equiv \sqrt{\tilde{g}^2 - 6\chi_1 -2 \chi_2} \,,  \label{gencoupcont}
\end{equation}
is a generalized version of the SU(2) group coupling constant, $H$ is the Hubble parameter, and a dot means a derivative with respect to cosmic time, the following autonomous dynamical system is obtained:
\begin{eqnarray}
&& x^2 + 2 x y + y^2 + 2 z^4 + (40 x^2 y^2 - 80 x y^3 + 40 y^4) \alpha_1 \nonumber \\
&& + (-8 x^2 y^2 + 8 y^4) \alpha_2 + (-216 x y^3 - 62 y^4) \alpha_3 \nonumber \\
&&+ (-480 x y^3 - 120 y^4) \alpha_4 + (32 x y^3 + 12 y^4) (3 \alpha_5 + \alpha_6) \nonumber \\
&&+ (-4 x^2 y^2 - 8 x y^3 - 4 y^4) (\chi_3 + \chi_4 + 3 \chi_5 + 6 \chi_6 + 2 \chi_7) = 1 \,, \label{fried}
\end{eqnarray}
\begin{eqnarray}
&& x^2 + 2 x y + y^2 + 2 z^4 + (440 x^2 y^2 + 80 \sqrt{2} p y^3 - 80 x y^3 - 520 y^4 + 160 y^4 \epsilon) \alpha_1 \nonumber \\
&& + (-40 x^2 y^2 - 8 \sqrt{2} p y^3 - 112 x y^3 - 56 y^4 + 16 y^4 \epsilon) \alpha_2 \nonumber \\
&& + (648 x^2 y^2 + 108 \sqrt{2} p y^3 + 496 x y^3 - 214 y^4 + 92 y^4 \epsilon) \alpha_3 \nonumber \\
&& + (1440 x^2 y^2 + 240 \sqrt{2} p y^3 + 960 x y^3 - 600 y^4 + 240 y^4 \epsilon) \alpha_4 \nonumber \\
&& + (-96 x^2 y^2 - 16 \sqrt{2} p y^3 - 96 x y^3 + 12 y^4 - 8 y^4 \epsilon) (3 \alpha_5 + \alpha_6) \nonumber \\
&& + (4 x^2 y^2 + 8 x y^3 + 4 y^4) (\chi_3 + \chi_4 + 3 \chi_5 + 6 \chi_6 + 2 \chi_7) = -3 + 2\epsilon \,, \label{accel}
\end{eqnarray}
\begin{eqnarray}
&& \frac{p}{\sqrt{2}} + 3 x + 2 y + \frac{4 z^4}{y} - y \epsilon + (40 x^2 y + 20 \sqrt{2} p y^2 + 120 x y^2 - 200 y^3 + 40 y^3 \epsilon) \alpha_1 \nonumber \\
&& + (-8 x^2 y - 4 \sqrt{2} p y^2 - 24 x y^2 - 16 y^3) \alpha_2 + (-200 y^3 + 108 y^3 \epsilon) \alpha_3 \nonumber \\
&& + (-480 y^3 + 240 y^3 \epsilon) \alpha_4 + (24 y^3 - 16 y^3 \epsilon) (3 \alpha_5 + \alpha_6) \nonumber \\
&& + (-4 x^2 y - 2 \sqrt{2} p y^2 - 12 x y^2 - 4 y^3 + 4 y^3 \epsilon) (\chi_3 + \chi_4 + 3 \chi_5 + 6 \chi_6 + 2 \chi_7) = 0 \,. \label{psifield}
\end{eqnarray}
In the above expressions, $p$ and $\epsilon$ are defined as
\begin{eqnarray}
p \equiv \frac{\ddot{\psi}}{m_P H^2} \,, \label{pdef} \\
\epsilon\equiv - \frac{\dot{H}}{H^2} \,, \label{epsilondef}
\end{eqnarray}
$\epsilon$ being the first slow-roll parameter.  It is worthwhile noting that Eqs. (\ref{fried}) and (\ref{accel}) are the Einstein field equations while Eq. (\ref{psifield}), which is not independent of Eqs. (\ref{fried}) and (\ref{accel}), is given by the contracted Bianchi identity.  Moreover, the role of the dimensionless parameters $\chi_1$ and $\chi_2$, as seen in Eq. (\ref{defdynsys}), is just to renormalize the group coupling constant $\tilde{g}$.  The equation (\ref{fried}) is the constraint equation that allows us to reduce the dimensionality of the system to two so that the phase space is given in terms of the variables $x$ and $y$.  In fact, this equation is exhausted so that the two independent gravitational field equations are now Eqs. (\ref{accel}) and (\ref{psifield}) where the variable $z$ is replaced by its expression in terms of the variables $x$ and $y$. Thus, having four independent variables, $x$, $y$, $p$, and $\epsilon$, and the two independent equations (\ref{accel}) and (\ref{psifield}), the dynamical system is closed with the equations
\begin{eqnarray}
x' &=& \frac{p}{\sqrt{2}} + \epsilon x \,, \label{xevolution} \\
y' &=& x \,, \label{yevolution}
\end{eqnarray}
where a prime denotes a derivative with respect to the e-folds number defined as $N \equiv \int H \ dt$.  The expressions in Eqs. (\ref{xevolution}) and (\ref{yevolution}) are easily obtained from the definitions in Eqs. (\ref{defdynsys}), (\ref{pdef}), and (\ref{epsilondef}).  For an ample discussion of the application of dynamical systems to cosmology, see Refs. \cite{dscbook1,dscbook2,Bahamonde:2017ize,Jarv:2021qpp}.

\section{The attractor straight line in phase space and inflation} \label{inflation}

Is it possible to have a de Sitter constant-roll inflationary period in the framework of the GSU2P according to the action in Eq. (\ref{action})?  In agreement with the definition of a constant-roll inflationary period \cite{Motohashi:2014ppa,Motohashi:2017vdc,Motohashi:2019tyj}, the single degree of freedom that describes the cosmic triad, i.e. $\psi$, must have an acceleration directly proportional to the product of its velocity with the Hubble parameter.  Thus, 
\begin{equation}
\ddot{\psi} = \frac{1}{\beta} H \dot{\psi} \,, \label{croll}
\end{equation}
where $\beta$ is a dimensionless parameter.  This equation can be easily integrated assuming the constant-roll inflationary period is de Sitter:
\begin{equation}
\psi = \beta \frac{\dot{\psi}}{H} + \gamma \,,
\end{equation}
where $\gamma$ is a dimensionful integration constant.  In terms of the dimensionless variables that define the dynamical system, the above equation reads
\begin{equation}
y = \beta x + \delta \,, \label{straightline}
\end{equation} 
where $\delta$ is a dimensionless constant.  This corresponds to a straight line in phase space.  

Does this straight line, as an integral curve in phase space, really exist?  Does it describe a de Sitter period? How long is this inflationary period?  Is this straight line an attractor?  These questions are addressed in the following subsections.

\subsection{Existence of the straight line integral curve as a de Sitter solution}

The first thing to do is to check if a solution of the form in Eq. (\ref{straightline}) is compatible with the dynamical system.  In fact, for large values of ${\rm max}(|x|,|y|)$, such a solution does exist and is unique:
\begin{eqnarray}
\beta &=& \frac{70 \alpha_1 - 6 \alpha_2 + 
  108 \alpha_3 + 240 \alpha_4 - 48 \alpha_5 - 
 16 \alpha_6 + \chi_3 + \chi_4 + 3 \chi_5 +
 6 \chi_6 + 2\chi_7}{
70 \alpha_1+ 8 \alpha_2 + 19 \alpha_3 + 60 \alpha_4 - \chi_3 - \chi_4 - 3 \chi_5 - 6 \chi_6 - 2\chi_7} \,, \label{beta} \\
\delta &=& 0 \,. \label{delta}
\end{eqnarray}
Moreover, the $\epsilon$ parameter vanishes along this line, so the latter represents a de Sitter inflationary period.  It is important to emphasize that {\it this solution exists only for large values of ${\rm max}(|x|,|y|)$}.

\begin{figure}%
    \centering
    \subfloat[\centering Primordial inflation \label{fig1a}]{{\includegraphics[width=9cm]{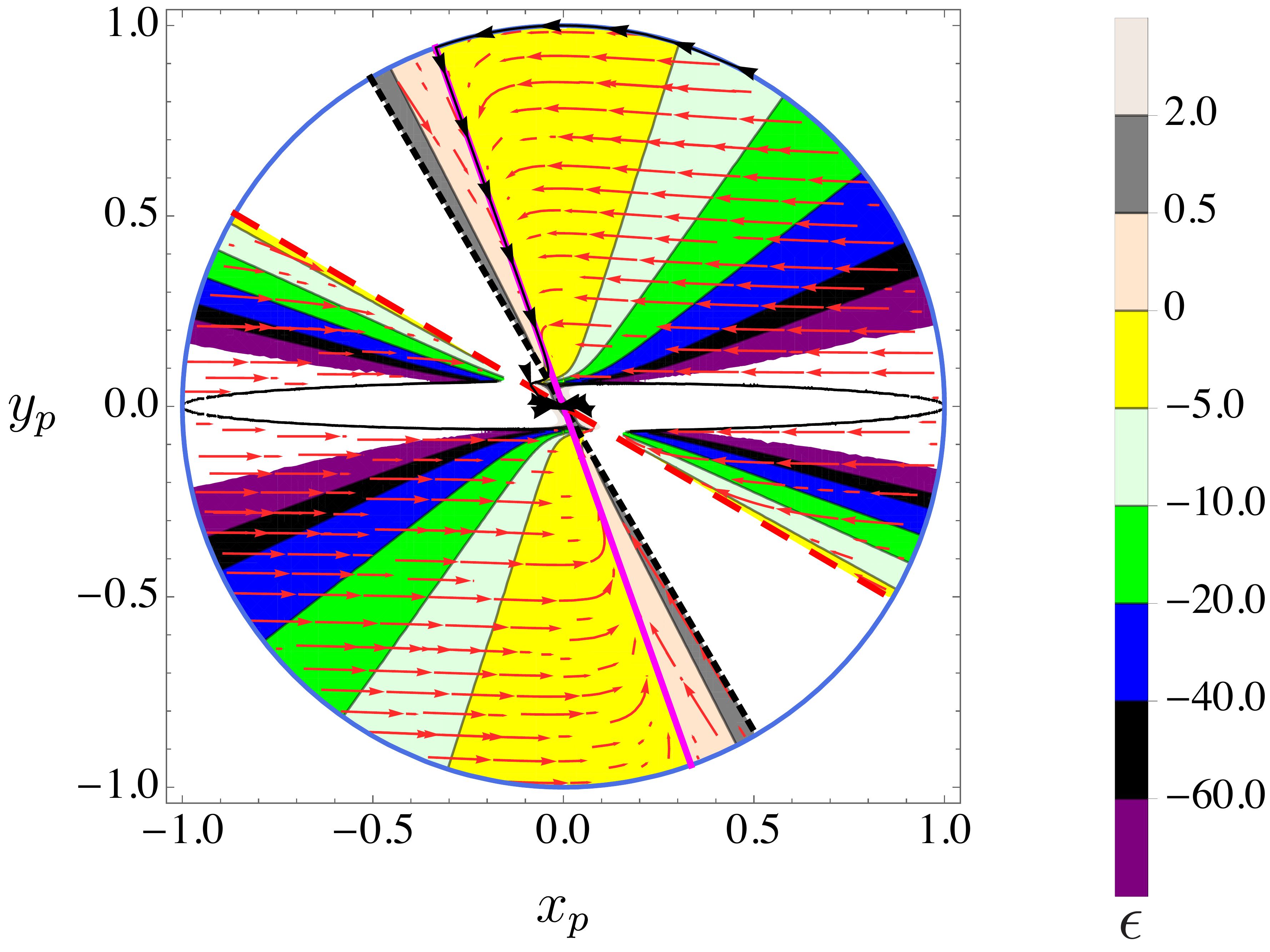} }}%
    \qquad
    \subfloat[\centering Late time inflation \label{fig1b}]{{\includegraphics[width=9cm]{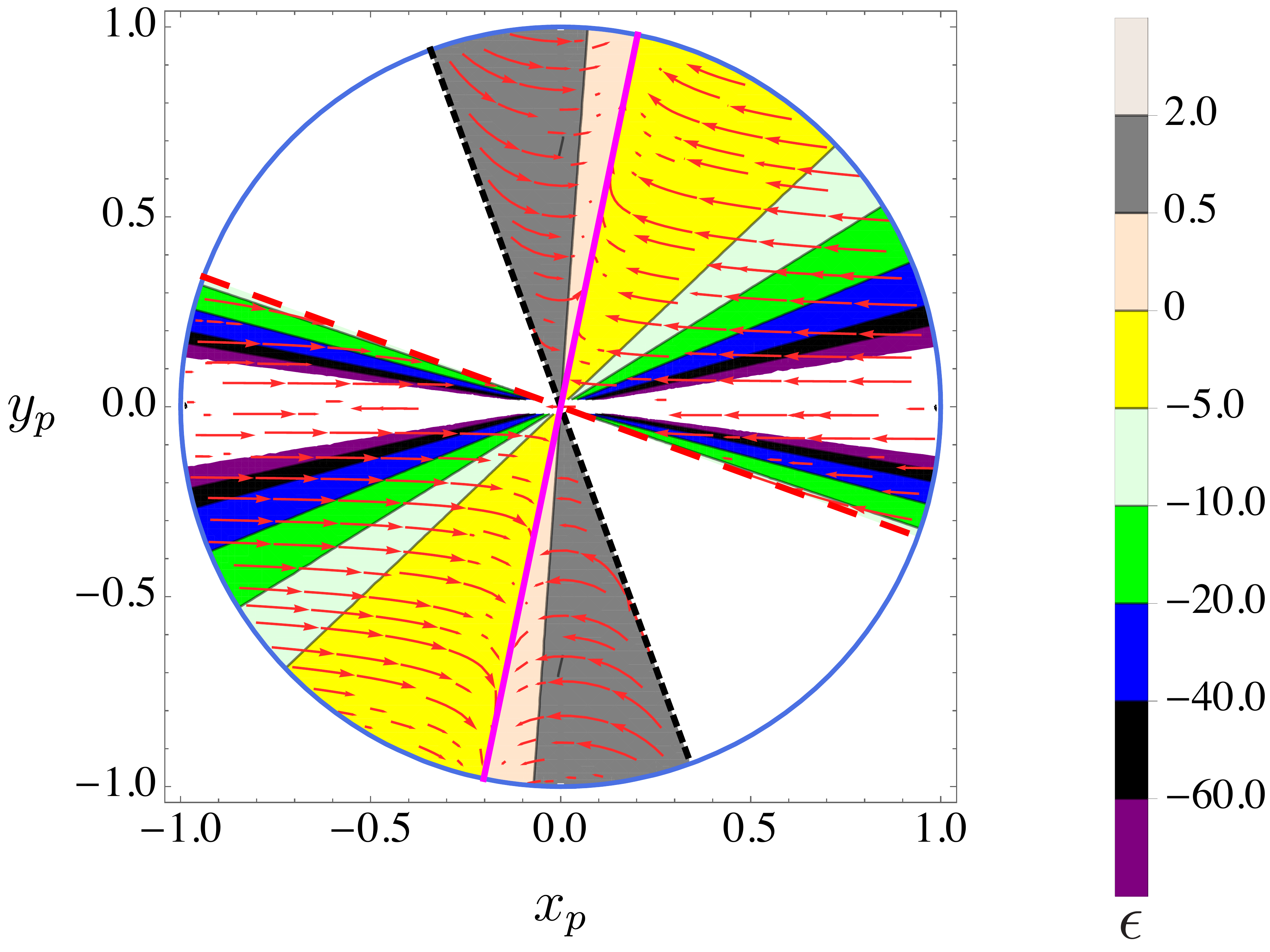} }}%
    \caption{Phase space portraits in the Poincar\'e circle for primordial and late time inflation in the GSU2P with the action of Eq. \ref{newaction}.  $\alpha_1 = 1, \alpha_3 = 1.0008$, and $\chi_5 = -1.965$ for Fig. \ref{fig1a} whereas $\alpha_1 = 1, \alpha_3 = 2$, and $\chi_5 = 8$ for Fig. \ref{fig1b}.  $\chi_4$ and $\chi_6$ may take any values. The zoom parameter in both plots is $\zeta = 25$.  The allowed regions are filled with the integral curves that describe the dynamical system flow.  Several contours for the first slow-roll parameter $\epsilon$ parameter are drawn and distinguished by a colour code.  The solid magenta straight line $y_p = \beta x_p$ corresponding to the de Sitter constant-roll inflation is also drawn.  Notice that, in both cases, the attraction basin of the constant-roll solution fills almost all the allowed region.  Notice as well that there exist other two straight lines in Fig. \ref{fig1a}, the dashed ones (black and red), each one being identified with part of the borders, at large max($|x|,|y|$), of the three connected but mutually disconnected regions whose union makes the whole allowed region.  Fig. \ref{fig1a} also portraits, with a solid arrowed black curve, a numerical solution to the system with initial conditions $x_{\rm ini} = 5 \times 10^9$ and $y_{\rm ini} = 10^{10}$. Primordial inflation ends up when the system reaches the central zone where the Y-M term begins dominating, giving way to damped oscillations of the $\psi$ field and to a radiation dominated period.}%
    \label{fig1}%
\end{figure}

Since the phase space is two dimensional, we can plot the allowed region together with the integral curves.  This is easier to do by projecting the whole $x-y$ plane onto the Poincar\'e circle through the following redefined variables:
\begin{equation}
x_p \equiv \frac{x}{\sqrt{\zeta^2 + x^2 + y^2}} \,, \hspace{5mm} y_p \equiv \frac{y}{\sqrt{\zeta^2 + x^2 + y^2}} \,,
\end{equation}
which can be inverted as
\begin{equation}
x \equiv \frac{\zeta x_p}{\sqrt{1 - x_p^2 - y_p^2}} \,, \hspace{5mm} y \equiv \frac{\zeta y_p}{\sqrt{1 - x_p^2 - y_p^2}} \,,
\end{equation}
where $x_p^2 + y_p^2  < 1$ and $\zeta$ is a positive real number that permits to zoom in and zoom out the central region of the phase space in the Poincar\'e circle.  The $x$ and $y$ axes are mapped onto the $x_p$ and $y_p$ axes respectively and the straight line $y = \beta x$ is mapped onto the straight line $y_p = \beta x_p$.  Figs. \ref{fig1a} and \ref{fig1b} portrait the phase space in the Poincar\'e circle for the parameters
\begin{eqnarray}
\alpha_2 &=& 2 \alpha_3 \,, \nonumber \\
\alpha_6 &=& -20 \alpha_1 + 6 \alpha_3 - 3 \alpha_5 \,, \nonumber \\
\chi_3 &=& 0 \,, \nonumber \\
\alpha_4 &=& -2 \alpha_1 + \frac{7}{20} \alpha_3 \,, \nonumber \\
\alpha_5 &=& -\frac{20 \alpha_1 - 14 \alpha_3}{3} \,, \nonumber \\
\chi_7 &=& 5 \alpha_1 + \alpha_3 - \frac{\chi_4}{2} - 3 \chi_6 \,, \label{parpres}
\end{eqnarray}
which lead to the action
\begin{eqnarray}
S = \int d^4 x \ \sqrt{-g} && \Big[\mathcal{L}_{E-H} + \mathcal{L}_{Y-M} + \chi_4 \left(\mathcal{L}_2^4 - \frac{\mathcal{L}_2^7}{2}\right) + \chi_5 \mathcal{L}_2^5 + \chi_6 \left( \mathcal{L}_2^6 - 3 \mathcal{L}_2^7 \right) \nonumber \\
&& + \alpha_1 \left(\mathcal{L}_{4,2}^1- 2 \mathcal{L}_{4,2}^4 - \frac{20}{3} \mathcal{L}_{4,2}^5 + 5 \mathcal{L}_2^7 \right) \nonumber \\
&& + \alpha_3 \left(2 \mathcal{L}_{4,2}^2 + \mathcal{L}_{4,2}^3 + \frac{7}{20} \mathcal{L}_{4,2}^4 + \frac{14}{3} \mathcal{L}_{4,2}^5 - 8 \mathcal{L}_{4,2}^6 + \mathcal{L}_2^7 \right)  \Big] \,. \label{newaction}
\end{eqnarray}
In Fig. \ref{fig1a}, $\alpha_1 = 1, \alpha_3 = 1.0008$, and $\chi_5 = -1.965$ have been chosen, leaving $\chi_4$ and $\chi_6$ free since they do not contribute to the field equations (\ref{fried}) to (\ref{psifield});  the zoom parameter in this plot is $\zeta = 25$.  Fig. \ref{fig1b} is plotted with the same parameters prescription as in Eq. (\ref{parpres}) and the same zoom but having chosen $\alpha_1 = 1, \alpha_3 = 2$, and $\chi_5 = 8$.  The reason for the different constraints among coupling constants in Eq. (\ref{parpres}) is that, this way, the perturbative tensor sector of the theory under study behaves as in GR at least up to second order so that there are neither ghosts nor Laplacian instabilities;  moreover, the gravity waves propagate at the light speed.  This is properly discussed in Sec. \ref{reduction}.

\begin{figure}%
    \centering
    \subfloat[\centering Primordial inflation \label{fig2a}]{{\includegraphics[width=9cm]{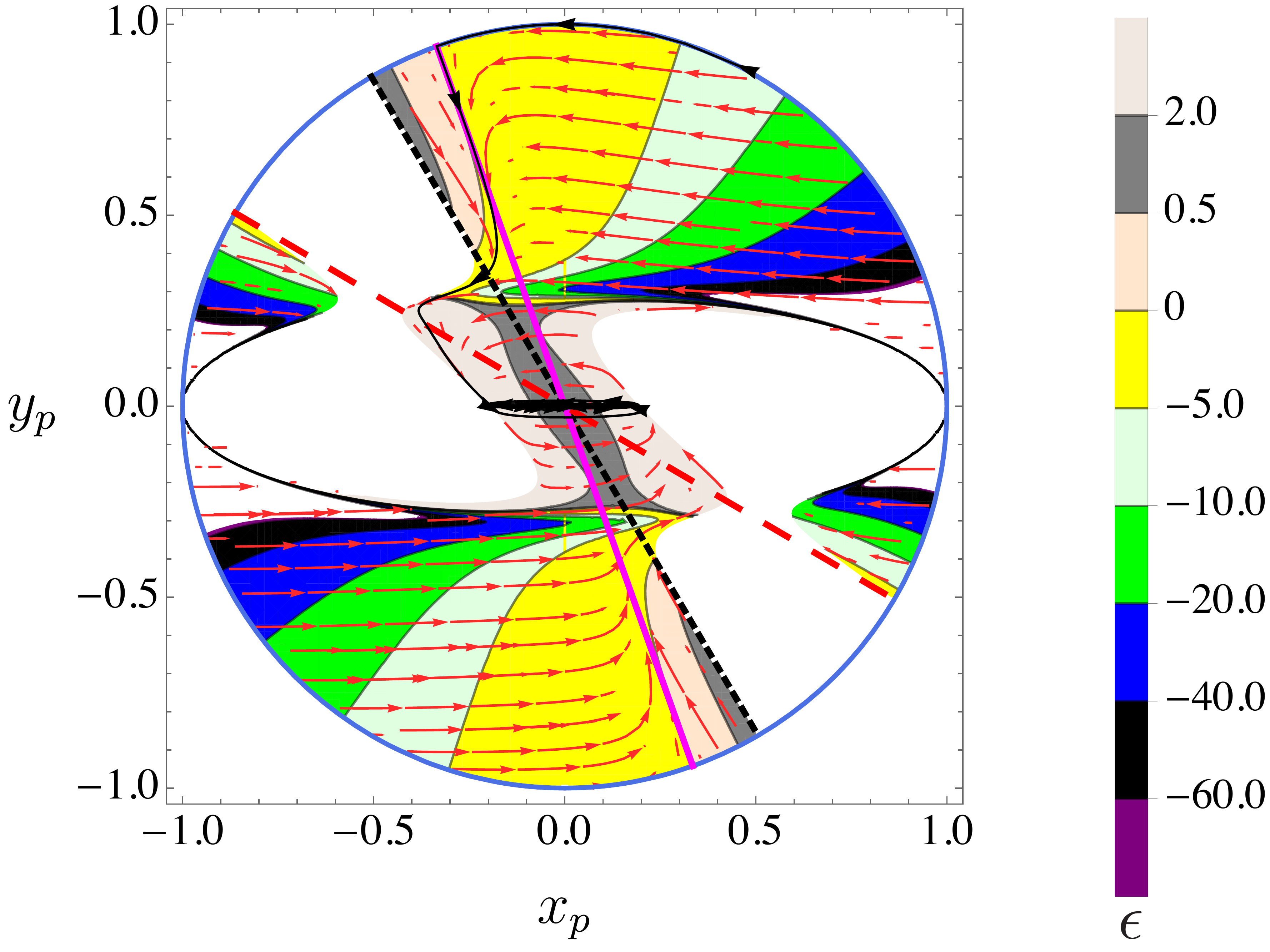} }}%
    \qquad
    \subfloat[\centering Primordial inflation \label{fig2b}]{{\includegraphics[width=9cm]{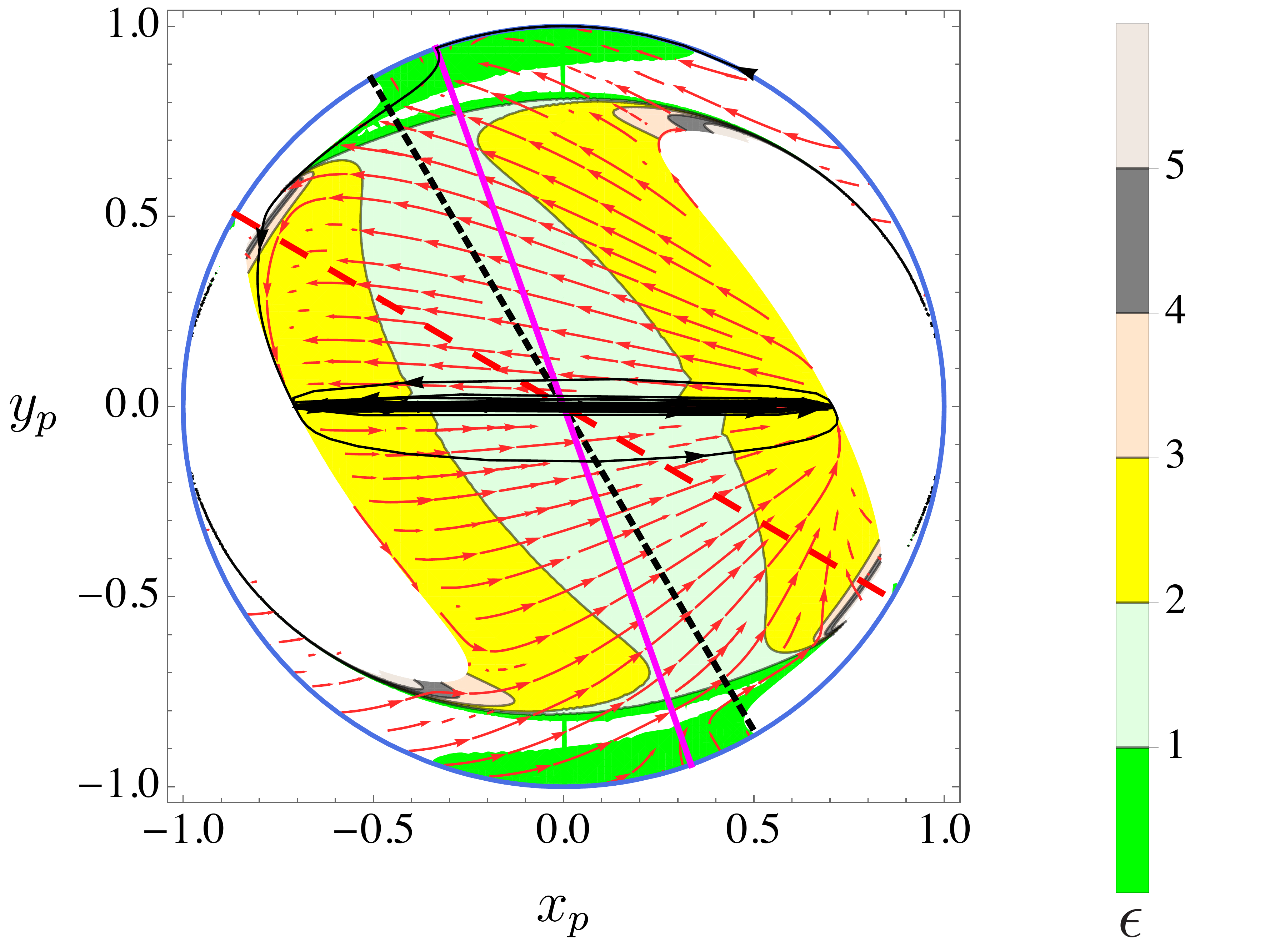} }}%
    \caption{Phase space portraits in the Poincar\'e circle corresponding to Fig. \ref{fig1a} with the central zone zoomed ($\zeta = 5$ for Fig. \ref{fig2a} and $\zeta = 1$ for Fig. \ref{fig2b}).  These plots reveal with more detail the end of primordial inflation when the system reaches the central zone of the plot ({\rm max}($|x|,|y|) \lesssim 1$).  In this stage of the evolution, the Y-M and the Einstein-Hilbert terms dominate, the vector fields experience damped oscillations, and the system enters into a radiation dominated period.}%
    \label{fig2}%
\end{figure}

The slope of the straight line is important since the negative slope is useful to describe primordial inflation whereas the positive slope is useful to describe dark energy.  The reason is that, since the straight line crosses the origin of the phase space, it will lie on the second and fourth quadrants if the slope is negative;  this means, in turn, that the evolution of the system along the straight line is such that it moves towards the origin, $|x|$ and $|y|$ then reducing as time passes (see Fig. \ref{fig1a}).  Eventually, $|x|$ and $|y|$ are small enough and the straight line is not a solution anymore.  The system will now be dominated by the lower powers of $\psi$, i.e., by the canonical kinetic (i.e., Y-M) term together with the Einstein-Hilbert term, and primordial inflation will have a graceful exit into a radiation dominated era (see Figs. \ref{fig1a} and \ref{fig2}).  In contrast, if the slope is positive, the straight line will lie on the first and third quadrants, which means that the system will evolve towards larger values of $|x|$ and $|y|$;  there will be no graceful exit into anything making this scenario very convenient to eventually describe the late inflationary period (see Fig. \ref{fig1b}).  The latter scenario was explored in Ref. \cite{Rodriguez:2017wkg}, only for a linear combination of $\mathcal{L}_{4,2}^1$ and $\mathcal{L}_{4,2}^4$, although without the presence of matter.  Indeed, not only $\mathcal{L}_{4,2}^1$ and $\mathcal{L}_{4,2}^4$ but also $\mathcal{L}_{4,2}^3$, by themselves, have the potential of describing dark energy.  In contrast, $\mathcal{L}_{4,2}^2$, it being the only purely non-Abelian term in the considered action, and $\mathcal{L}_2^3$ to $\mathcal{L}_2^7$, by themselves, have the potential to describe primordial inflation;  however, the set $\mathcal{L}_2^3$ to $\mathcal{L}_2^7$ is unable to produce $\epsilon = 0$ unless at least one of the $\mathcal{L}_{4,2}$ Lagrangian pieces is present.  Here is where the power of the generalized Proca terms built with the symmetric version of the gauge field strength tensor in its Abelian version, the $S$ terms as they were called in Ref. \cite{Rodriguez:2017wkg}, resides.  On the other hand, the Lagrangian pieces $\mathcal{L}_{4,2}^5$ and $\mathcal{L}_{4,2}^6$, by themselves, are unable to provide the constant-roll solution of Eq. (\ref{straightline}).

\subsection{Amount of inflation}

\begin{figure}%
    \centering
    \subfloat[\centering  \label{fig3a}]{{\includegraphics[width=7cm]{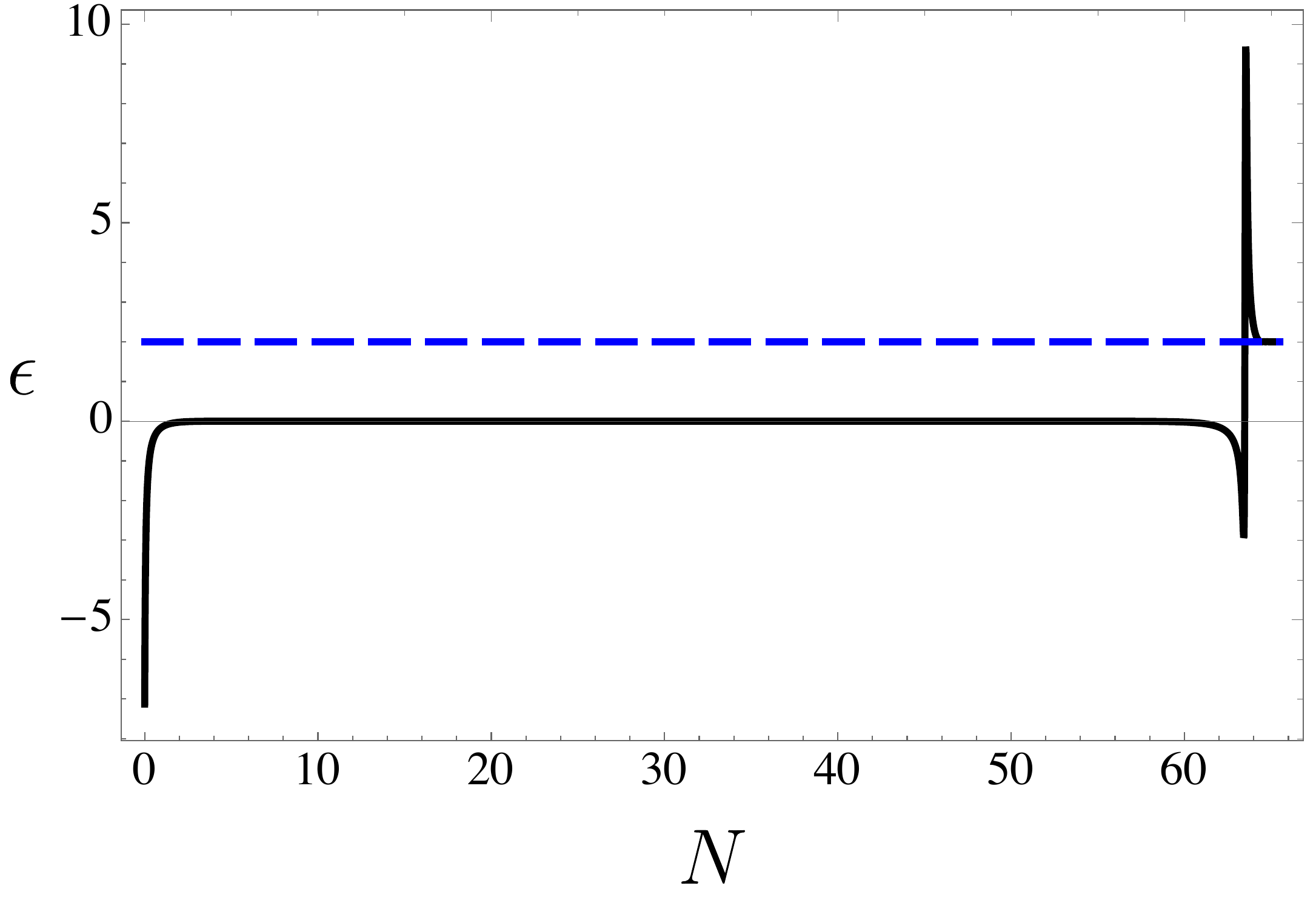} }}%
    \qquad
    \subfloat[\centering  \label{fig3b}]{{\includegraphics[width=7cm]{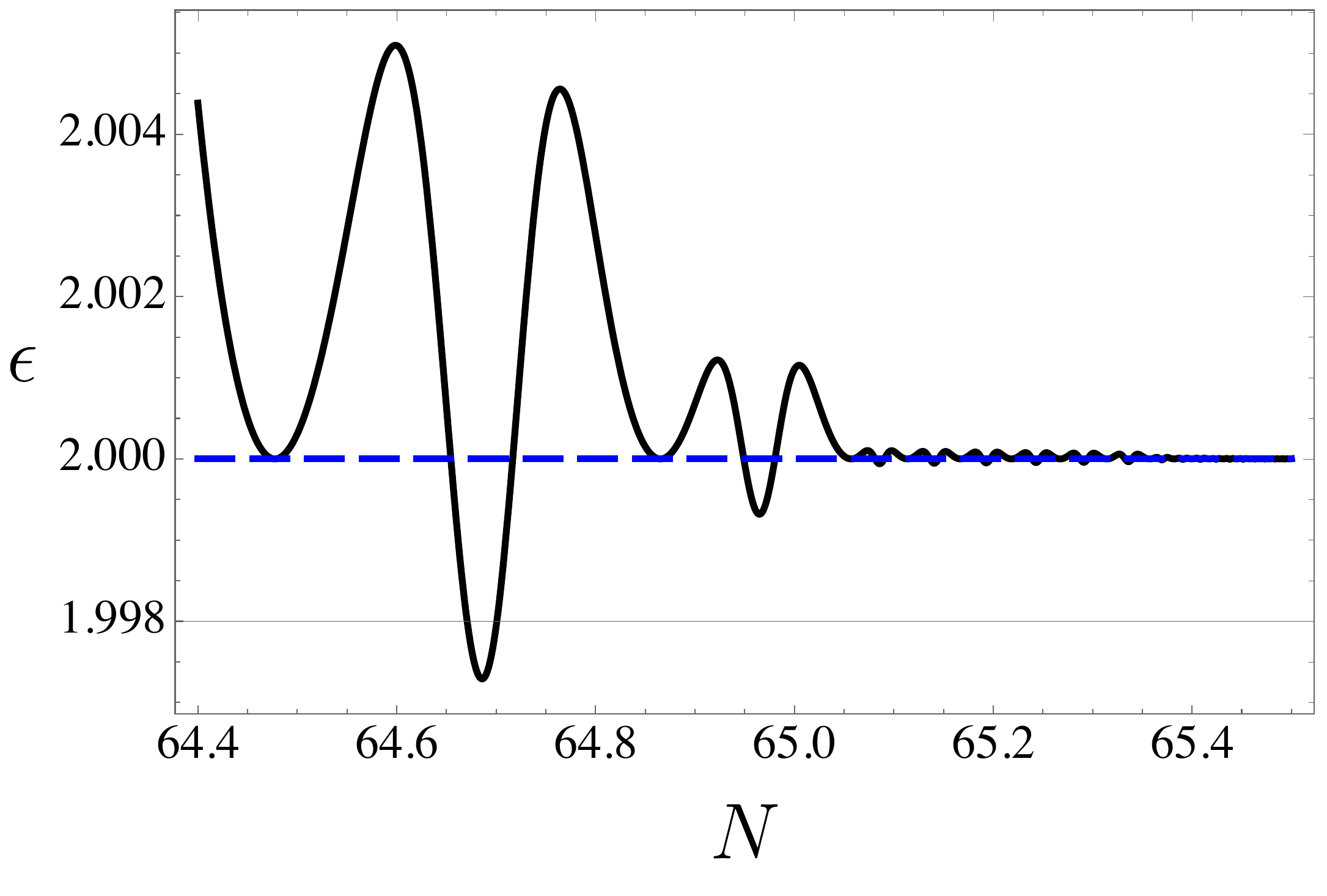} }}%
    \caption{The first slow-roll parameter $\epsilon$ (in solid black) vs the amount of inflation $N$ for the primordial inflation scenario of Figs. \ref{fig1a} and \ref{fig2}.  As can be seen in Fig. \ref{fig3a}, the $\epsilon$ parameter quickly approaches zero and stays there for a number of e-folds that is wholly consistent with Eq. (\ref{efolds}).  This is enough to solve the classical problems of the standard cosmology.  As seen in Fig. \ref{fig3b}, the $\epsilon$ parameter starts oscillating around 2 (dashed blue line) after the end of inflation, signaling a radiation dominated period powered by the canonical kinetic (i.e. Y-M) term.}%
    \label{fig3}%
\end{figure}

To successfully solve the three classic problems of the standard cosmology, the flatness, horizon, and unwanted relics problems \cite{Guth:1980zm}, the primordial inflationary period must last at least some 60 to 70 e-folds \cite{Lyth:2009zz,Dimopoulosbook,Peter:2013avv,Weinberg:2008zzc}.  For the scenario under study, the amount of inflation or e-folds number is very easy to calculate:
\begin{equation}
N \equiv \int H \ dt = \int H \frac{d \psi}{\dot{\psi}} = \int \frac{d y}{x} \,,
\end{equation}
which, along the straight line $y = \beta x$, reduces to
\begin{equation}
N = \int^{y_{\rm fin}}_{y_{\rm ini}} \beta \frac{d y}{y} \approx - \beta \ln{|y_{ \rm ini}|} \,, \label{efolds}
\end{equation}
where $y_{\rm ini}$ is the initial value of the $y$ variable and the final value $y_{\rm fin} \sim \mathcal{O}(1)$ is given by the transition to the canonical kinetic (i.e., Y-M) term dominated era.  Several numerical runs have demonstrated that Eq. (\ref{efolds}) is an excellent approximation;  an example of this is given by Figure \ref{fig3} where the $\epsilon$ parameter vs $N$ is shown for the parameters employed to plot Figs. (\ref{fig1a}) and (\ref{fig2}).

\begin{figure}%
    \centering
    \subfloat[\centering Primordial inflation \label{fig4a}]{{\includegraphics[width=9cm]{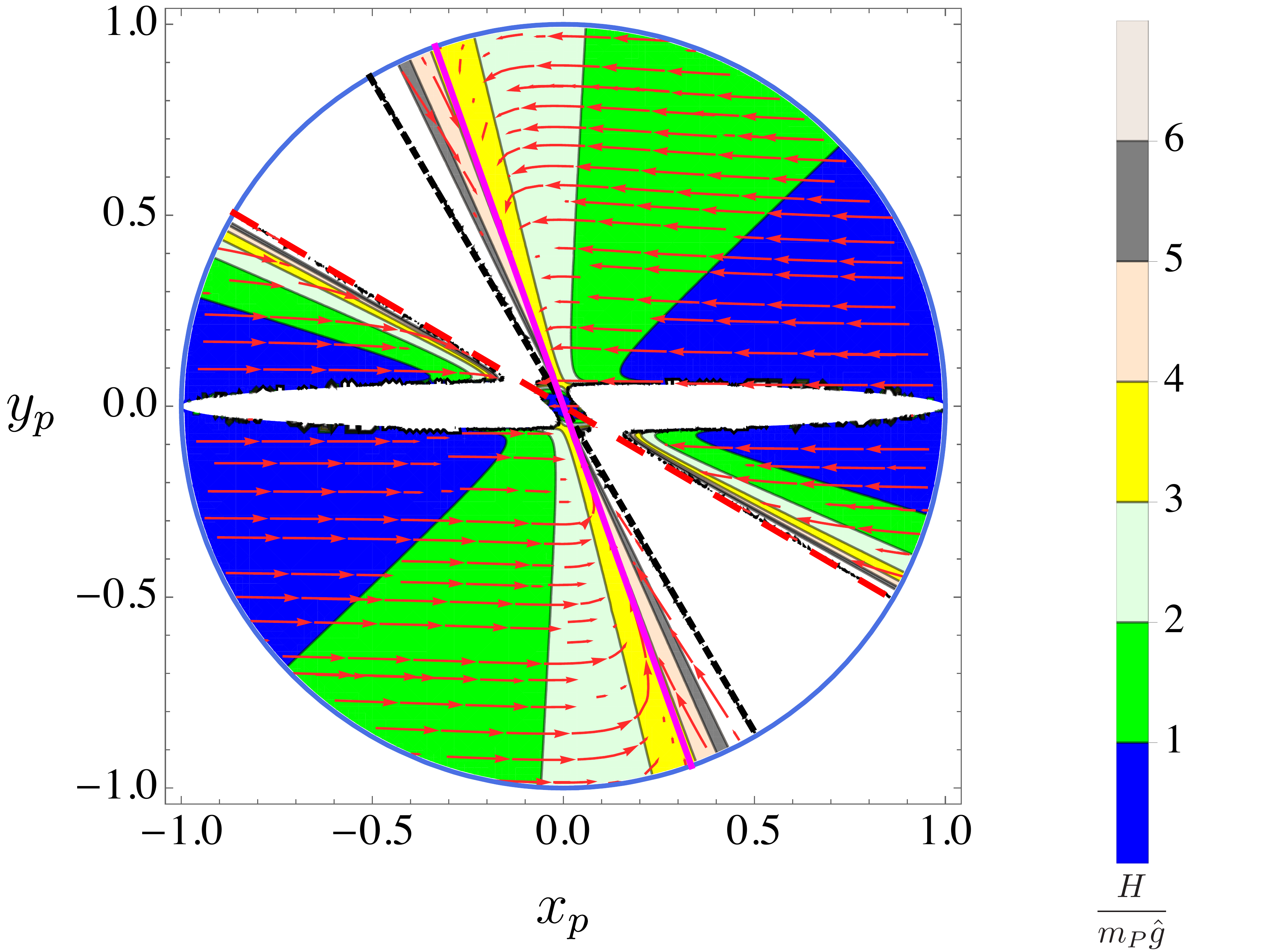} }}%
    \qquad
    \subfloat[\centering Primordial inflation \label{fig4b}]{{\includegraphics[width=9cm]{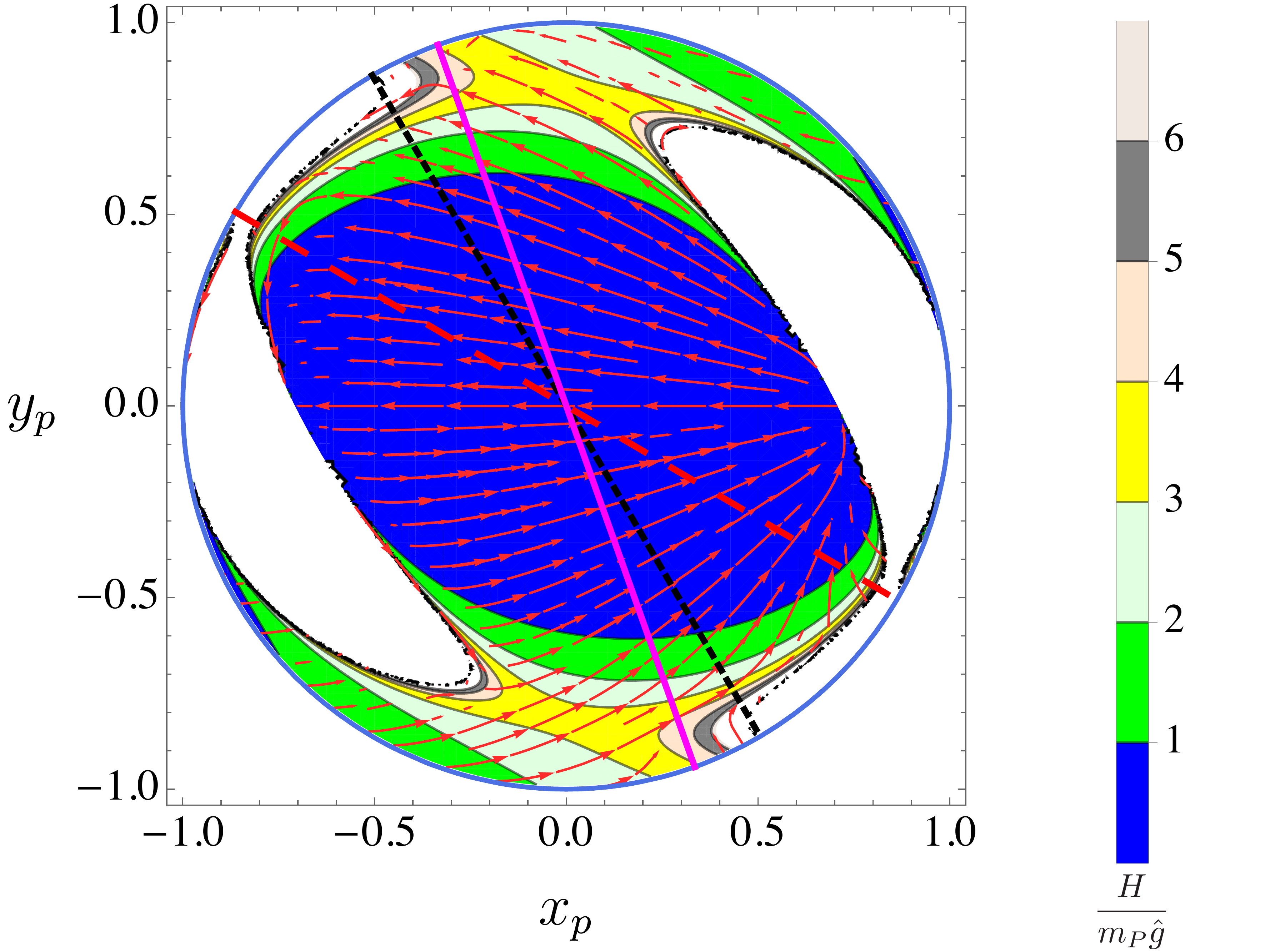} }}%
    \caption{Contours of the Hubble parameter in units of $m_P \hat{g}$ for the primordial inflation scenario of Figs. \ref{fig1a} and \ref{fig2}.  The zoom parameters for Figs. \ref{fig4a} and \ref{fig4b} are $\zeta = 25$ and $\zeta = 1$, respectively.  Any value for $H$ can actually be obtained by tuning the generalized SU(2) group coupling constant $\hat{g}$.}%
    \label{fig4}%
\end{figure}

Thus, in order to have $N \gtrsim 70$, $y_{\rm ini}$ must satisfy $|y_{\rm ini}| \gtrsim e^{70/|\beta|}$, i.e.,
\begin{equation}
\frac{\psi_{\rm ini}}{m_P} \gtrsim \sqrt{2} \ e^{70/|\beta|} \,.
\end{equation}
It is quite easy, thereby, to require a super Planckian initial value for $\psi$;  however, what matters is the energy scale during inflation given by $H$.  According to the definitions in Eq. (\ref{defdynsys}), 
\begin{equation}
\frac{H}{m_P} = \hat{g} \ \frac{y^2}{z^2} \,,
\end{equation}
where $z$ can be written in terms of $x$ and $y$ thanks to the constraint in Eq. (\ref{fried}).  Therefore, the energy scale of primordial inflation in this scenario can be tuned to the desired value just by choosing the value of the generalized version of the SU(2) group coupling constant given in Eq. (\ref{gencoupcont}).  Fig. \ref{fig4} displays contours of $H/(m_P \hat{g})$ in the Poincar\'e circle for the primordial inflation scenario of Figs. (\ref{fig1a}) and (\ref{fig2}).

\subsection{Attractor condition}

The scenario depicted in this section where a straight line in phase space has been found, portraying a constant-roll de Sitter inflationary period, is predictive only if this straight line is an attractor.  A little perturbation in the $x$ variable, for $y$ unperturbed, must result in the system coming back to the straight line.  For a given $y$ along the straight line, let's call it $y_s$, the evolution is given by the Eq. (\ref{yevolution}), i.e., by
\begin{equation}
\frac{y_s'}{x_s'} = \frac{x_s}{x_s'} \,,
\end{equation}
where $x_s$ is the respective value along the straight line.  A little perturbation $\delta x = x - x_s$, for $y$ unperturbed, follows the evolution given by
\begin{eqnarray}
\frac{y'}{x'} &=& \frac{x}{x'(x,y)} \nonumber \\
&=& \frac{x_s + \delta x}{x_s' (x_s) + \delta x' (x_s, \delta x)} \nonumber \\
&\approx& \frac{x_s}{x_s'(x_s)} + \frac{1}{x_s'(x_s)} \left[ \delta x - \frac{x_s \ \delta x' (x_s,\delta x)}{x_s'(x_s)} \right] \,,
\end{eqnarray}
at first order in the perturbations.  Considering the second quadrant, if the straight line is to be an attractor, 
\begin{equation}
\frac{y'}{x'} \mathop{\lessgtr} \frac{y_s'}{x_s'} \hspace{5mm} {\rm for} \hspace{5mm} \delta x \mathop{\gtrless} 0 \,,
\end{equation}
i.e.,
\begin{equation}
\delta x \mathop{\lessgtr} \frac{x_s \ \delta x' (x_s,\delta x)}{x_s'(x_s)} \hspace{5mm} {\rm for} \hspace{5mm} \delta x \mathop{\gtrless} 0 \,.
\end{equation}
The previous expression is valid as long as $x_s'(x_s) > 0$, which is indeed satisfied as can be observed by taking into account Eqs. (\ref{defdynsys}), (\ref{pdef}), (\ref{xevolution}), and (\ref{croll}).  Therefore, no matter which the sign of $\delta x$ is, the attractor condition reads
\begin{equation}
\frac{x_s}{x_s'(x_s)} \frac{\partial x'}{\partial x} \Big|_{x_s} > 1 \,. \label{attcondition}
\end{equation}
This condition constrains the possible values that the coupling constants in the action may take.  It is, however, not the only constraint.  The next sections will explore this and other constraints, some strictly necessary and some imposed for technical easiness, that significantly reduce the available parameter space.

The attractor condition in Eq. (\ref{attcondition}) can also be employed to maximize the basin of attraction.  Under the requirement that the straight line described by Eqs. (\ref{straightline}) - (\ref{delta}) represents primordial inflation (i.e., it must have negative slope) and that, of course, is inside the allowed region in phase space, the basin of attraction is maximized when the straight line that works as border, for large max($|x|,|y|$), of the allowed connected region in phase space that encloses the constant-roll straight line (see the tiny-dashed black straight line in, e.g., Fig. \ref{fig1}) becomes a repeller.  The opposite condition to that in Eq. (\ref{attcondition}) must then be applied to such border straight line.

\section{The perturbative tensor sector and the reduction in the number of free parameters} \label{reduction}

Fourteen free parameters, $\tilde{g}$, the seven $\alpha_i$, and the six $\chi_i$, are too much to analyze the constraints on them that will be presented later.  It is easier, therefore, to notoriously reduce the number of free parameters by imposing additional constraints that are convenient but not completely necessary.  An example of the latter is to request the perturbative tensor sector of the theory to behave as in GR, at least up to some order in perturbation theory.

Previous analysis about the stability conditions necessary for avoiding ghost and Laplacian instabilities in the perturbative tensor sector of the  GSU2P, see Ref. \cite{Gomez:2019tbj}, are extended here.  An alternative route is however adopted in this paper by choosing the coupling constants such that the second-order action of the theory is reduced to that of GR, inheriting in this way the luminal nature of the propagation speeds and releasing the theory of instabilities. For this class of vector-tensor theories, we have two types of tensor modes:  the one related to gravitational waves plus a tensor mode associated to the vector fields. Assuming a FLRW background, the tensor perturbation of the metric tensor reads
\begin{equation}
\delta g_{ij}=a^{2}(t)h_{ij} \,,
\end{equation}
while that of the vector field reads
\begin{equation}
\delta A_{i}^{a}=a(t)t_{i}^{a} \,,
\end{equation}
where both $h_{ij}$ and $t_{i}^{a}$ are symmetric and satisfy the transverseness and tracelessness conditions 
\begin{eqnarray}
\partial^{i}{h_{ij}}=h_{i}^{i} &=& 0 \,, \nonumber \\
\delta_{a}^{i}\partial_{i}t_{j}^{a}=\delta_{a}^{i}t_{i}^{a} &=& 0 \,. 
\end{eqnarray}
We define the polarization states $+$ and $\times$ for the metric and vector field tensor perturbations in the circular basis as follows:
\begin{eqnarray}
   \delta g_{11} &=&-\delta g_{22}=a^{2}h_{+} \,, \hspace{1cm} \delta g_{12}=a^{2}h_{\times} \,, \nonumber \\
    \delta A_{\mu}^{1} &=& a(0,t_{+},t_{\times},0,0) \,, \hspace{1cm} \delta A_{\mu}^{2}=a(0,t_{\times},-t_{+},0,0) \,,
\end{eqnarray}
which are oriented, without loss of generality, so that the perturbations propagate along the $z$-axis direction.  After expanding each piece of the action in Eq. (\ref{action}) up to second order in perturbations  and integrating by parts as appropriate,  the quadratic action containing the kinetic $K$,  the Laplacian $L$ and the mixed (containing products of first-order time and spatial derivatives) $C$ matrices can be written in a compact way \cite{BeltranJimenez:2017cbn} as
  \begin{equation}
    S_{T}^{2}=\int d^{3}x\; dt\; \left(a^{3} \dot{\vec{x}}^{T} K\; \dot{\vec{x}}  +a\;\partial\vec{x}^{T} L\; \partial\vec{x}+a^{2}\dot{\vec{x}}^{T} C \, \partial\vec{x} \right),
\end{equation}
where $\vec{x}^{T} \equiv (m_{P} h_{+}, t_{+},m_{P} h_{\times},t_{\times})$ and the dimension of the matrices are  determined by the number of degrees of freedom. The non-vanishing components of the kinetic matrix are
\begin{eqnarray}
    K_{11}&=&K_{33}=\frac{1}{8} \left[2+\psi^{4} (160\alpha_{1}+31\alpha_{3}+60 \alpha_{4}+6\alpha_{5}+10 \alpha_{6}) \right]\,, \nonumber \\
    K_{22}&=&K_{44}=1+2 \psi^2 (10 \alpha_{1}+\alpha_{2}-\chi_{3}-\chi_{4}-6\chi_{6}-2\chi_{7}) \,, \nonumber \\
    K_{12}&=&K_{21}=K_{34}=K_{43} = -\psi^3 (20 \alpha_{1}+\alpha_{2}+3 \alpha_{5}+2 \alpha_{6}) \,,
\end{eqnarray}
while those of the Laplacian matrix are
\begin{eqnarray}
    L_{11}&=&L_{33}=-\frac{1}{8} \left[2+\psi^{4} (240\alpha_{1}+7\alpha_{3}+60 \alpha_{4}+18\alpha_{5}+14 \alpha_{6}) \right]\,, \nonumber \\
    L_{22}&=&L_{44}=-\left[1+2 \psi^2 (10 \alpha_{1}+\alpha_{2}-2\chi_{3}-\chi_{4}-6\chi_{6}-2\chi_{7})\right] \,, \nonumber \\
    L_{12}&=&L_{21}=L_{34}=L_{43}= \psi^3 (20 \alpha_{1}+2\alpha_{2}-2\alpha_{3} +3\alpha_{5}+2 \alpha_{6}) \,.
    \end{eqnarray}
Off-diagonal elements of $K$ and $L$ correspond purely to mixed tensor modes.  Regarding $C$, there are no contributions to it since  parity-violating terms have not been included in the action of Eq. (\ref{action}) as they are incompatible at the background level with the spherical symmetry;  however, if they had been included, their respective coupling constants $\tilde{\alpha}_i$ would have had to be set to zero in order to make their contributions to the $C$ matrix vanish in agreement with the aim of reproducing GR. Thus, deviations of GR plus Y-M theory are sourced by self interactions of the vector field and non-minimal couplings to gravity. If they are not handled with care, instabilities may arise. 

In practice, building up a ghost-free theory demands both a positive definite kinetic matrix and positive squared propagation speeds.  In the case of GR, including also the standard Y-M term, stability is ensured by construction.  Such a procedure illuminates us to do the following.  Instead of calculating the cumbersome eigenvalues of $K$ and the even more cumbersome squared propagation speeds $c_t$, the latter by employing the expression 
\begin{equation}
\det(c_{t}^{2}K-c_{t}C+L)=0 \,,
\end{equation} 
and next imposing the corresponding conditions over the coupling constants as commonly done, a shorter and safer way to leave the GSU2P free of instabilities can be followed just by turning off the terms that deviate from GR plus the Y-M term, i.e., by making 
\begin{equation}
K = -L = K_{\rm GR} \,,  \label{perGRcond}
\end{equation} 
where $K_{\rm GR}$ is the kinetic matrix in GR. Another advantage of this procedure is the resulting background-$\psi$-independent constraints on the free parameters, in contrast to those found, for instance, in Refs. \cite{DeFelice:2016yws,Gomez:2019tbj}. The condition in Eq. (\ref{perGRcond}) leads to the constraints in Eq. (\ref{parpres}) which translates into a considerable reduction of the free parameters.  The first three constraints result from requiring the solutions for $c_t$ to be equal to 1, i.e., $K = -L$;  the other three constraints result from requiring $K = K_{\rm GR}$.  

This is one possibility among others much less restrictive.  Since the matrices now take the plain form characteristic of GR, one may think mistakenly of having eliminated the genuine effects of the GSU2P at the perturbative level, but this is not the case at all. Its effects on the tensor perturbations have just turned off for the purpose of stability but those on the scalar and vector perturbations remain. In this regard, the cosmological implications for scalar and vector perturbations have not been screened yet so the latter must be calculated in order to check their healthiness and to confront the primordial inflation scenario presented in this paper with the cosmic microwave background and large-scale structure observations.

\section{Graceful exit of inflation} \label{gracexit}

\begin{figure}%
    \centering
    \subfloat[\centering Failed primordial inflation \label{fig5a}]{{\includegraphics[width=9cm]{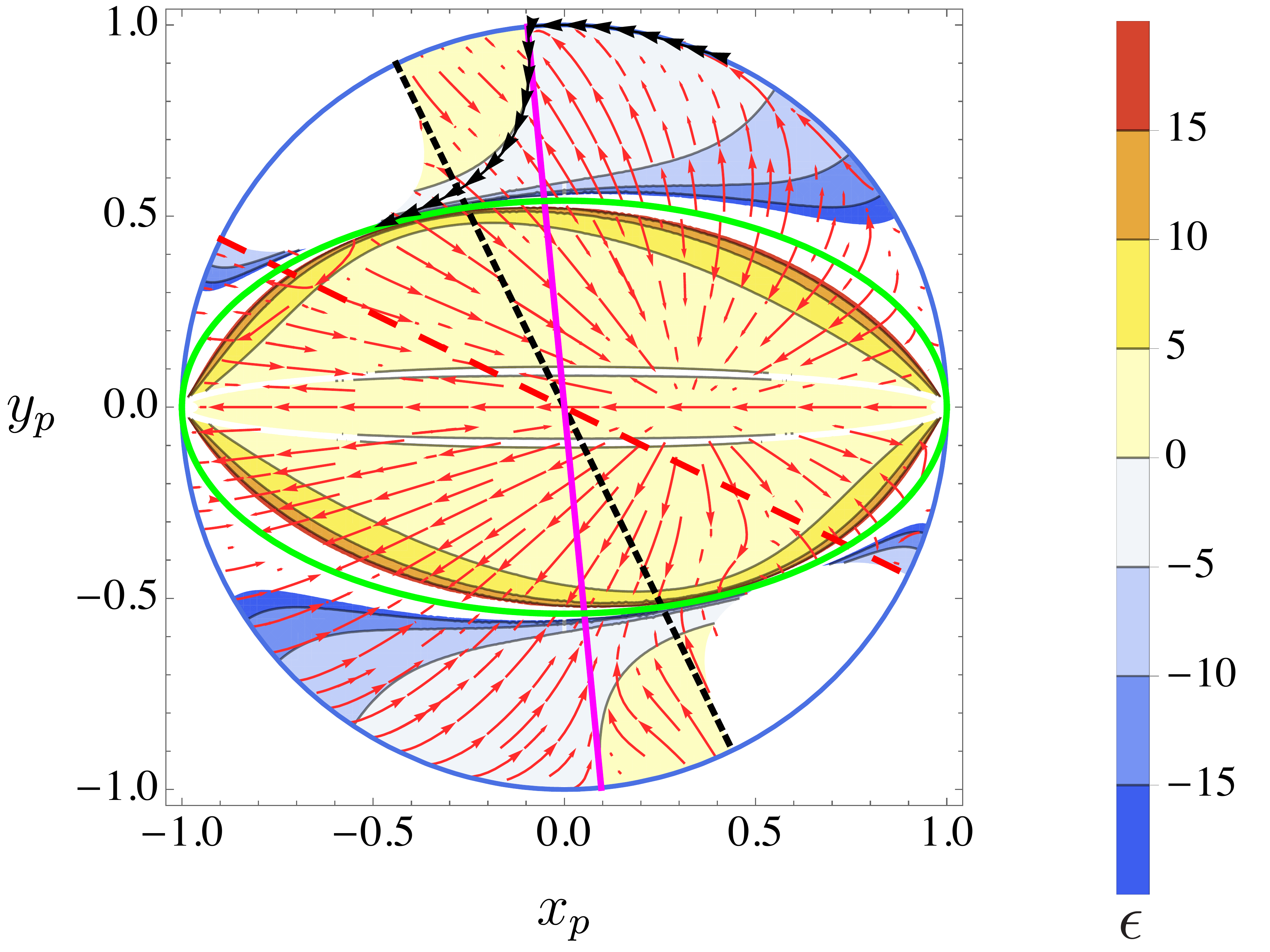} }}%
    \qquad
    \subfloat[\centering Failed primordial inflation \label{fig5b}]{{\includegraphics[width=9cm]{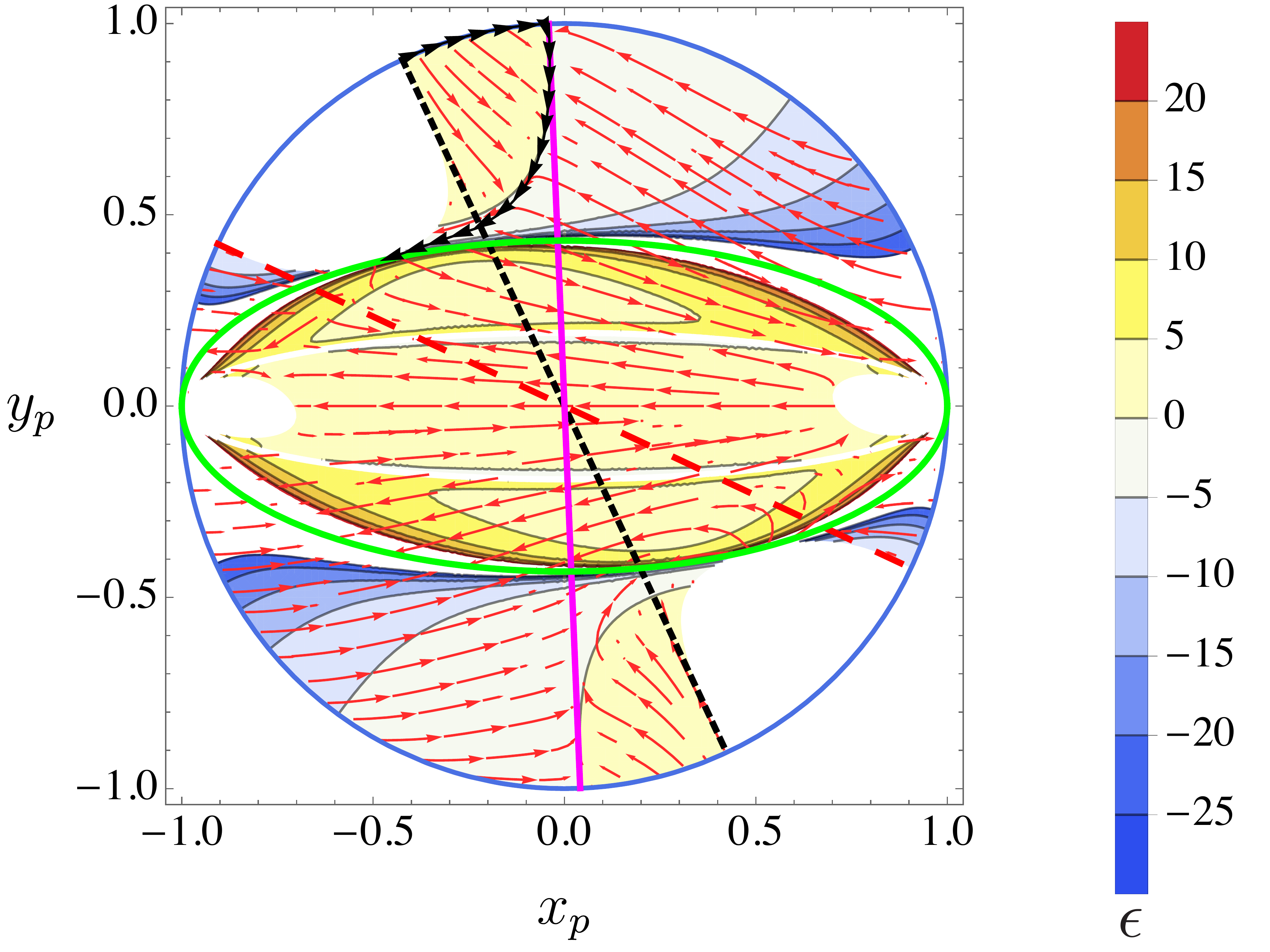} }}%
    \caption{Singularities can be met if the free parameters of the model are not properly chosen.  These singularities correspond to either $|x'|= |\epsilon| = \infty$ or even undefined, splitting the allowed region of phase space where the constant-roll straight line lies into three connected but mutually disconnected regions.  The borders between these regions are shown in the plots as the elliptic green and white stripes.  Fig. \ref{fig5a} corresponds to the parameters $\alpha_1=1, \alpha_3=5, \chi_5=90$ with a zoom parameter $\zeta = 0.3$;  the numerical solution in black arrows has initial conditions $x_{\rm ini} = 4 \times 10^9, y_{\rm ini} = 10^{10}$.  Fig. \ref{fig5b} corresponds to the parameters $\alpha_1=1, \alpha_3=1.1, \chi_5=0$ with a zoom parameter $\zeta = 1$;  the numerical solution in black arrows has initial conditions $x_{\rm ini} = -4 \times 10^9, y_{\rm ini} = 10^{10}$. In both plots, contours of the first slow-roll parameter $\epsilon$ are drawn and distinguished by a colour code.  The numerical solutions stop when reaching the first singularity, avoiding the passage to the central zone and, thereby, precluding the graceful exit of inflation into the radiation dominated stage.}%
    \label{fig5}%
\end{figure}

The scenario under study has a built-in mechanism to terminate inflation and continue with a radiation dominated period.  As has been already discussed, and leaving aside the Einstein-Hilbert term, for high values of max($|x|,|y|$), the dynamics is dominated by the terms involving four powers of the vector fields, i.e., by the self-interaction piece of the Y-M term plus the other GSU2P Lagrangian pieces in the action of Eq. (\ref{action}).  In contrast, for ${\rm max}(|x|,|y|) \lesssim 1$, what has been called before ``the central zone'' of the phase space, the dynamics is dominated by the non-self-interacting piece of the Y-M term.  This allows the end of primordial inflation as well as the onset of a radiation dominated period characterized by quick oscillations of the $\epsilon$ parameter around 2 as can be seen, for instance, in Figs. \ref{fig2} and \ref{fig3b}.  In order to access the central zone, the quantities that describe the evolution of the system must not meet singularities.  For the case at hand, the singularities are produced whenever the denominator in
\begin{eqnarray}
x' &=& \Big\{-2 + 2 x^2 + 3 x y + 2 y^2 + 
   4 y^2 \Big[-1600 x (7 x - 6 y) y^4 \alpha_1^2 + 
      16 y^3 (-129 x^3 - 877 x^2 y + 573 x y^2 + 81 y^3) \alpha_3^2 \nonumber \\
&&  + 
      3 \chi_5 (-x^2 + x (-3 + x^2) y + 3 (-1 + x^2) y^2 + 3 x y^3 + 
         y^4 - 12 y^3 (x + y)^3 \chi_5) + 
      2 \alpha_3 (53 x^3 y + 3 y^2 (7 - 9 y^2) \nonumber \\
&&      - 3 x y (23 + 7 y^2) + 
         x^2 (-3 + 39 y^2) - 
         24 y^3 (23 x^3 + 34 x^2 y + 9 x y^2 - 12 y^3) \chi_5) + 
      20 y \alpha_1 (-5 x^3 - 3 x^2 y \nonumber \\
&&      + 3 y (-1 + y^2) + 
         3 x (2 + y^2) + 
         4 y^2 (2 (12 x^3 + 155 x^2 y - 120 x y^2 - 9 y^3) \alpha_3 + 
            3 (4 x^3 + 5 x^2 y - 3 y^3) \chi_5))\Big]\Big\} \nonumber \\
&&            /\Big\{y + 
   160 y^5 (\alpha_1 - \alpha_3) \Big[1 + 
      4 y^2 (10 \alpha_1 - 16 \alpha_3 - 3 \chi_5)\Big] - 
   12 y^3 (2 \alpha_3 + \chi_5)\Big\} \,,  \label{largexp}
\end{eqnarray}
which, by the way, is the same denominator in $\epsilon$, becomes zero.  The previous expression is the same as that in Eq. (\ref{xevolution}) when $p$ and $\epsilon$ have been solved in terms of the variables $x$ and $y$.  Fig. \ref{fig5} presents the phase-space portraits with singularities for two different sets of free parameters together with particular numerical solutions.  The parameters in Fig. \ref{fig5a} are $\alpha_1=1, \alpha_3=5, \chi_5=90$ with a zoom parameter $\zeta = 0.3$ while the respective numerical solution has initial conditions $x_{\rm ini} = 4 \times 10^9, y_{\rm ini} = 10^{10}$.  In contrast, the parameters in Fig. \ref{fig5b} are $\alpha_1=1, \alpha_3=1.1, \chi_5=0$ with a zoom parameter $\zeta = 1$ while the respective numerical solution has initial conditions $x_{\rm ini} = -4 \times 10^9, y_{\rm ini} = 10^{10}$. The region of phase space where the constant-roll primordial inflationary solution lies is split into three connected but mutually disconnected regions whose common borders are the sets of points where the singularity just described is met.  As seen, the numerical solutions stop when the first singularity is reached so that the graceful exit of inflation into the radiation dominated stage fails.

To avoid such fate, the strategy is first to consider the case when the denominator is positive for large values of $|y|$ and to find out its absolute minimum requiring it to become strictly greater than zero.  Afterwards, the case when the denominator is negative for large values of $|y|$ must be considered and its absolute maximum must be found out requiring it to become strictly less than zero.  The joint constraints are presented in the next section.

\section{Available parameter space}  \label{avalparspace}

There exist four constraints that define the available parameter space for the primordial inflation scenario:
\begin{enumerate}
\item The slope of the straight line in Eq. (\ref{beta}) must be negative. \label{i1}
\item The straight line must be enclosed into the allowed phase space in the $x-y$ plane that is delineated by the constraint in Eq. (\ref{fried}) and the fact that $z^4 \geq 0$. 
\item The straight line must be an attractor, so Eq. (\ref{attcondition}) must be satisfied. \label{i3}
\item The system must be able to access the central zone, i.e., no singularities in $x'$ (given by Eq. (\ref{largexp})), and therefore in $\epsilon$, must exist. \label{i4}
\end{enumerate}

An optional constraint comes from the requirement of maximizing the attraction basin.  Thus, the straight line that serves as part of the border of the allowed region where the constant-roll straight line resides (see the tiny dashed black line in all the figures of primordial inflation) must be a repeller so the opposite condition to Eq. (\ref{attcondition}) must be applied to it.

\begin{figure}%
    \centering
    \subfloat[\centering Available parameter space - Curve \label{fig6a}]{{\includegraphics[width=7cm]{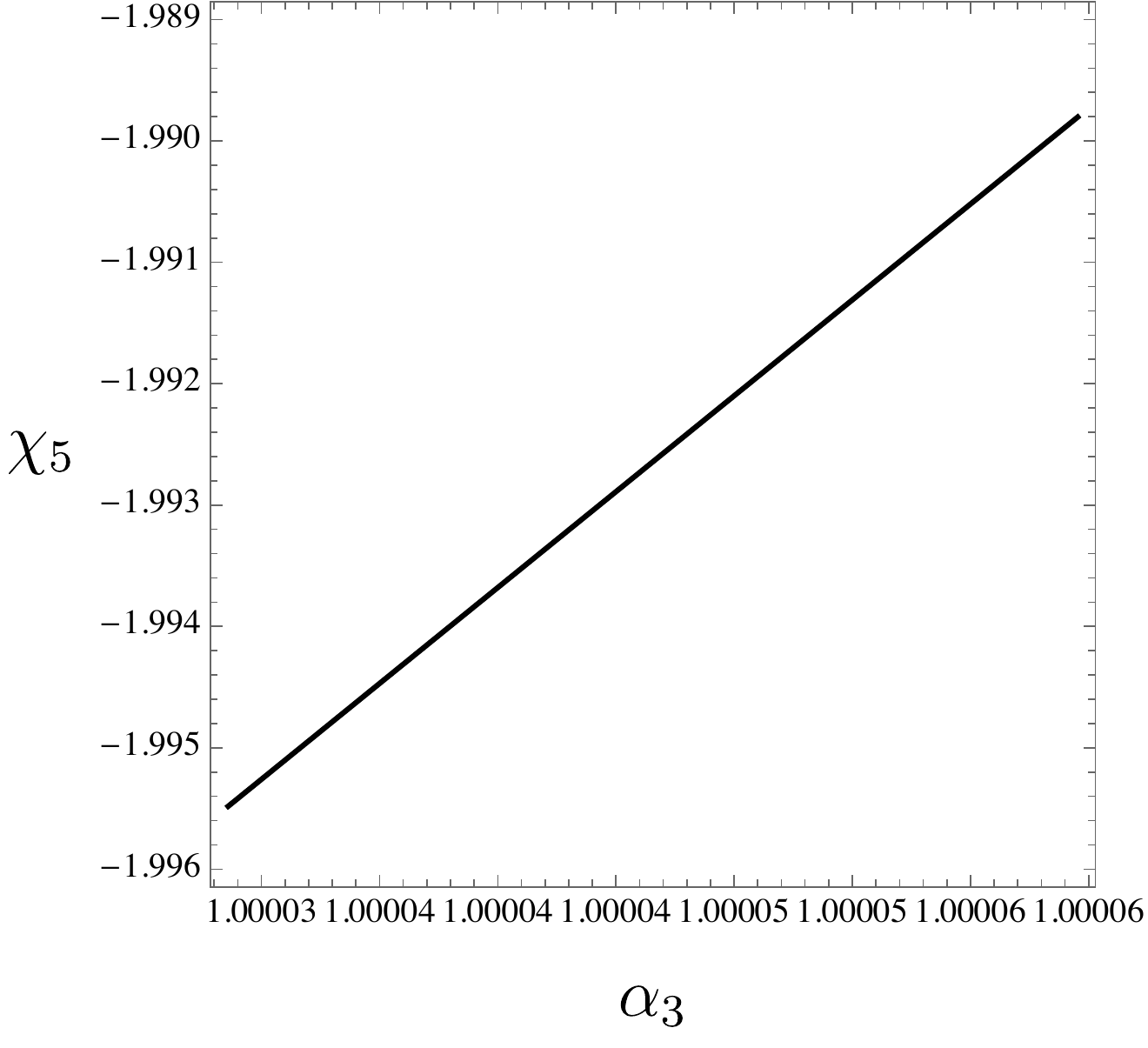} }}%
    \qquad  \hspace{1cm}
    \subfloat[\centering Available parameter space - Region I \label{fig6b}]{{\includegraphics[width=9cm]{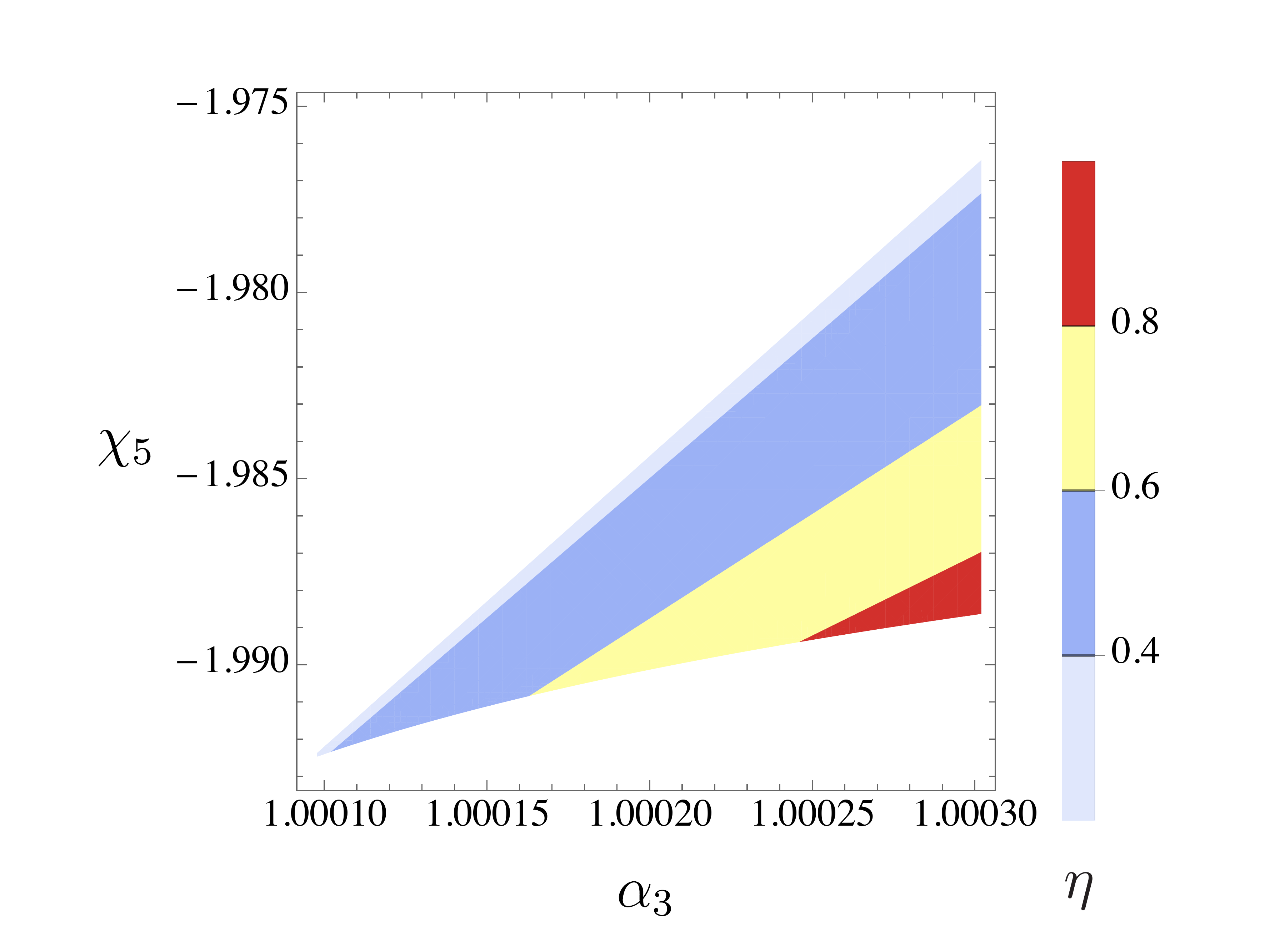} }}%
    \\
    \subfloat[\centering Available parameter space - Region II \label{fig6c}]{{\includegraphics[width=9cm]{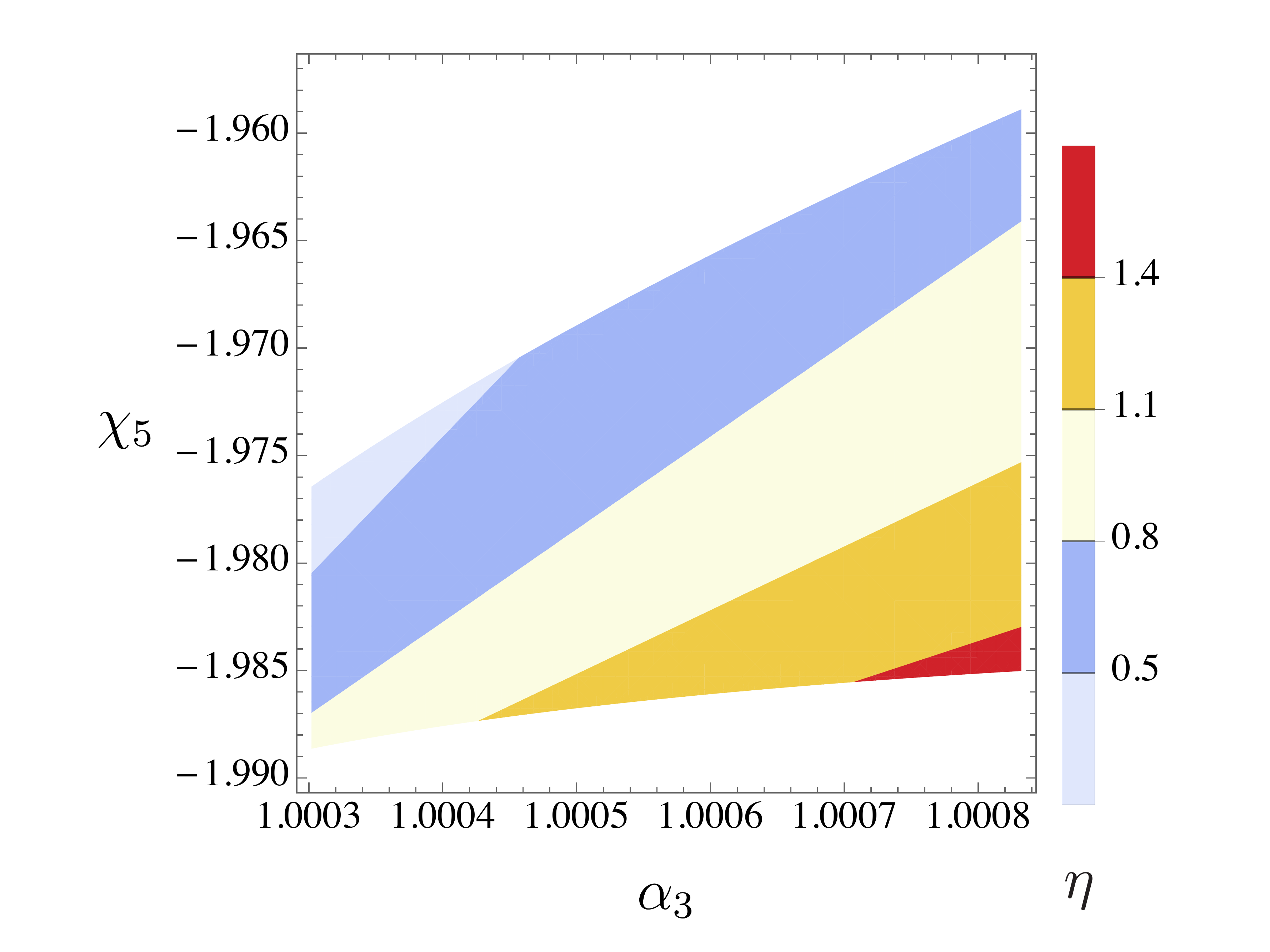} }}%
    \qquad
    \subfloat[\centering Available parameter space - Region III \label{fig6d}]{{\includegraphics[width=9cm]{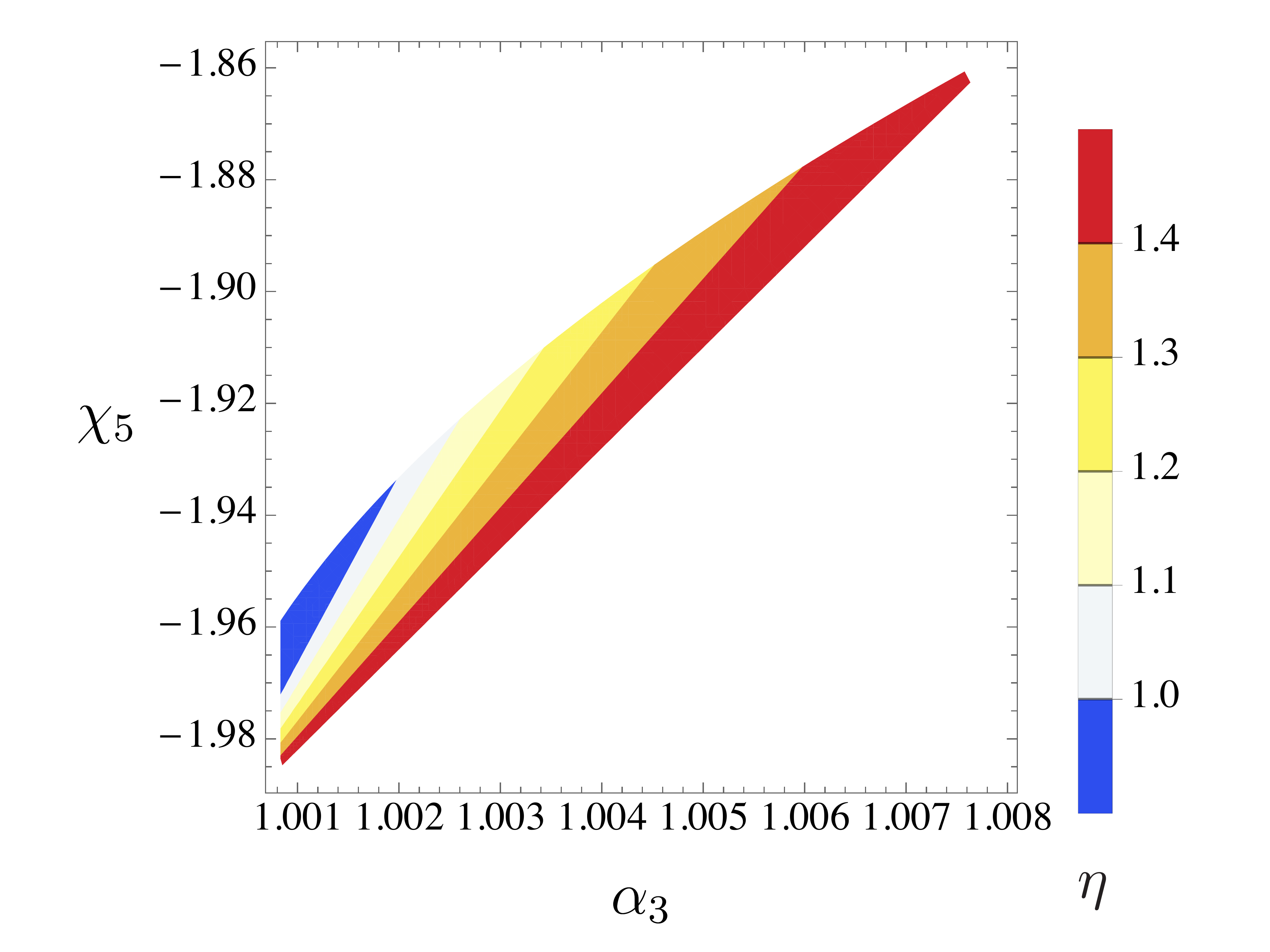} }}%
    \caption{The available parameter space split into one curve and three disjoint regions in the plane $\alpha_3 - \chi_5$ for $\alpha_1 = 1$.  Contours of the second slow-roll parameter $\eta$ are drawn and distinguished by a colour code ($\eta = 0$ in the curve of Fig. \ref{fig6a}).  The union of the curve and the three regions is shown in Fig. \ref{fig7}.}%
    \label{fig6}%
\end{figure}

\begin{figure}%
    \centering
    {\includegraphics[width=9cm]{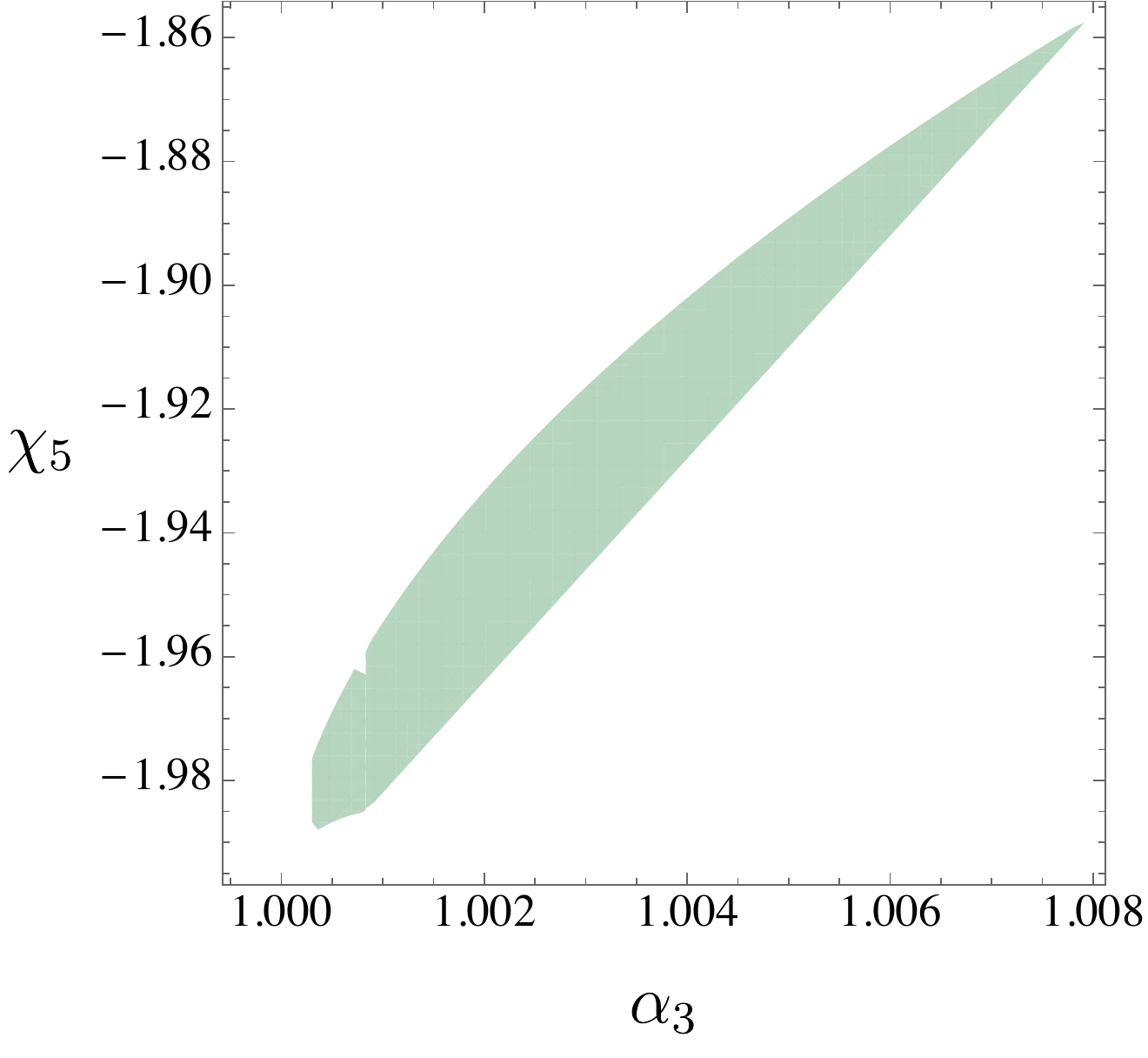} }%
    \caption{The union of the curve and the three disjoint regions of Fig. \ref{fig6} that represents the whole available parameter space for $\alpha_1 = 1$ in the plane $\alpha_3 - \chi_5$.}%
    \label{fig7}%
\end{figure}

Conditions \ref{i1} to \ref{i3} are satisfied as long as
\begin{eqnarray}
\alpha_3 > \alpha_1 \hspace{5mm} &\wedge& \hspace{5mm} -20 \alpha_1 + 18 \alpha_3 \leq \chi_5 \leq -160 \alpha_1 + 158 \alpha_3  \,, \nonumber \\
&\vee& \nonumber \\
\alpha_3 = \alpha_1 \hspace{5mm} &\wedge& \hspace{5mm} \chi_5 \neq -160 \alpha_1 + 158 \alpha_3 \,, \nonumber \\
&\vee& \nonumber \\
\alpha_3 < \alpha_1 \hspace{5mm} &\wedge& \hspace{5mm} \left(\chi_5 \leq -160 \alpha_1 + 158 \alpha_3 \hspace{5mm} \vee \hspace{5mm} \chi_5 > \frac{80 \alpha_1 - 86 \alpha_3}{3}\right) \,.  \label{inequal}
\end{eqnarray}
If the optional constraint is implemented, the inequalities in Eq. (\ref{inequal}) are strengthened to
\begin{equation}
\alpha_3 > \alpha_1 \hspace{5mm} \wedge \hspace{5mm} (-20 \alpha_1 + 18 \alpha_3 \leq \chi_5 \leq -80 \alpha_1 + 78 \alpha_3 \hspace{5mm} \vee \hspace{5mm} \chi_5 = -160 \alpha_1 + 158 \alpha_3) \,. \label{windowok}
\end{equation}
However, what reduces the most the available parameter space is the condition \ref{i4}.  It leaves no room except for the union of a curve and three small regions that are plotted in Figs. \ref{fig6} and \ref{fig7} in a very small interval around $\alpha_1 = 1$.  It is worth recalling at this point that the scenario presented in this paper is just a realization and, therefore, a proof of concept of what can be achieved when the full parameter space, with its fourteen free parameters or with general coupling functions in the most general case, is explored.  More free parameters or coupling functions always implies less stringent conditions so the potential to satisfy the five constraints enumerated in this section, together with the absence of ghost and Laplacian instabilities, for an ample zone of the parameter space is certainly high.

\section{Constant-roll inflation:  varieties and past singularities}  \label{crolls}

Having implemented the parameter reduction described in Sec. \ref{reduction}, the constant-roll de Sitter solution is represented by the straight line in phase space
\begin{equation}
y = \frac{80 \alpha_1 - 86 \alpha_3 - 3 \chi_5}{60 \alpha_1 - 54 \alpha_3 + 3 \chi_5} \ x \,,
\end{equation}
whose attraction basin covers most of the allowed region in phase space as can be seen in any of the preceding figures portraying successful primordial inflation.  This section discusses two interesting aspects of this scenario:  the well known varieties of inflation, slow roll and ultra slow roll, and the question of the existence of past singularities.

\subsection{Varieties of inflation}

\subsubsection{Slow-roll inflation}

Slow-roll inflation has been the archetypical model of inflation since its inception \cite{Lyth:2009zz,Dimopoulosbook,Peter:2013avv,Weinberg:2008zzc,Linde:1981mu,Linde:1982uu,Albrecht:1982wi,Liddle:1993fq}.  In the slow-roll approximation, the absolute value of the $\epsilon$ parameter defined in Eq. (\ref{epsilondef}) as well as the absolute value of the parameter
\begin{equation}
\eta \equiv 2\epsilon - \frac{\dot{\epsilon}}{2 H \epsilon} \,,
\end{equation}
are considered very small compared to one.  There exists an infinite tower of ordered dimensionless slow-roll parameters, each one describing the time evolution of the preceding one, while the first one, $\epsilon$, describes the fractional evolution of the Hubble parameter.  For most of the applications, the first and second slow-roll parameters, $\epsilon$ and $\eta$, are enough to describe the inflationary dynamics.  Indeed, it is easier to have a prolonged period of primordial inflation, long enough to solve the classical problems of the standard cosmology, if the slow-roll conditions are satisfied.  Moreover, consistency with the experimental data about the spectral index of the primordial curvature perturbations usually requires slow roll \cite{Lyth:2009zz,Dimopoulosbook}.

\begin{figure}%
    \centering
    \subfloat[\centering Slow-roll inflation \label{fig8a}]{{\includegraphics[width=9cm]{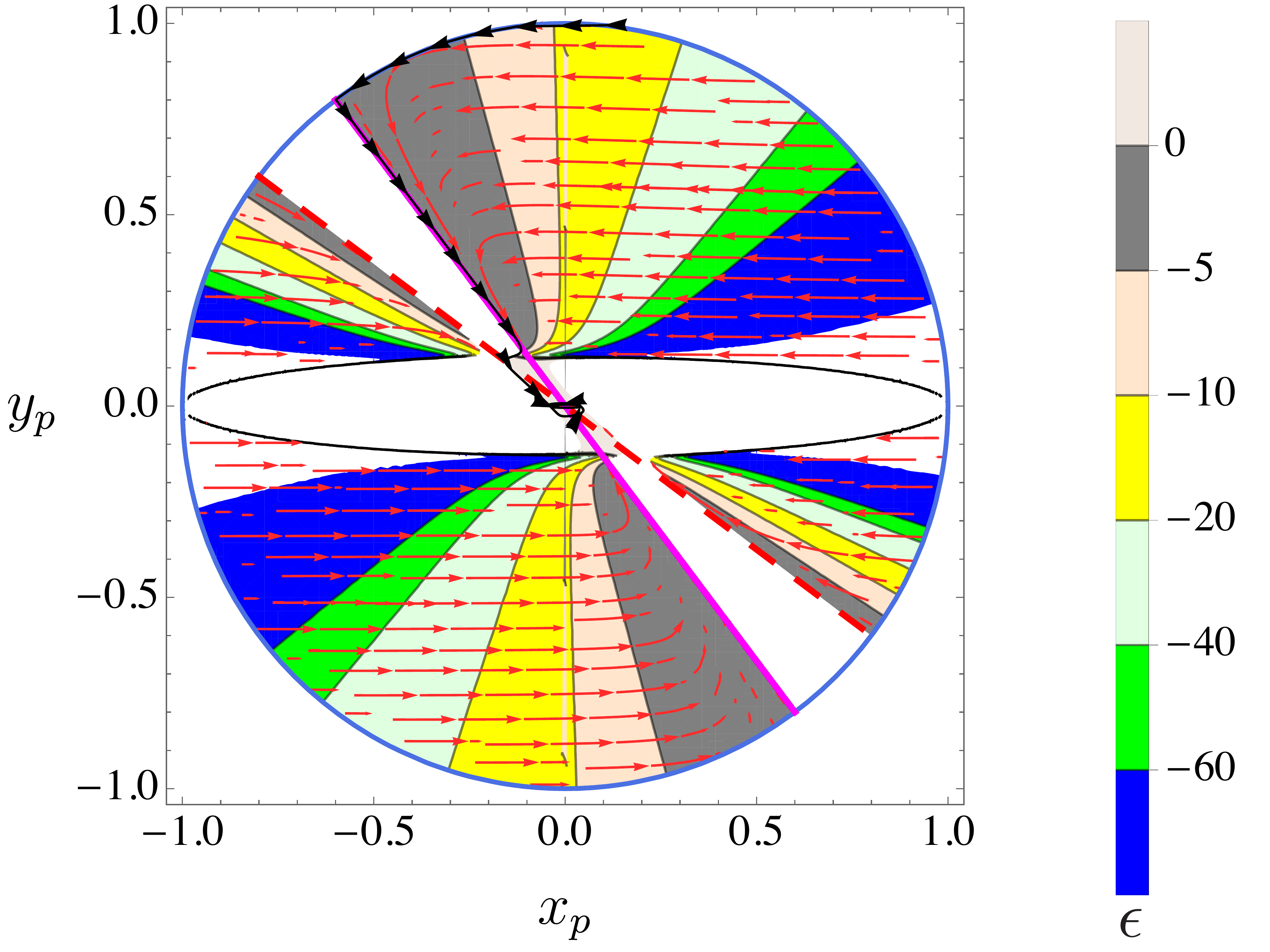} }}%
    \qquad
    \subfloat[\centering Slow-roll inflation \label{fig8b}]{{\includegraphics[width=9cm]{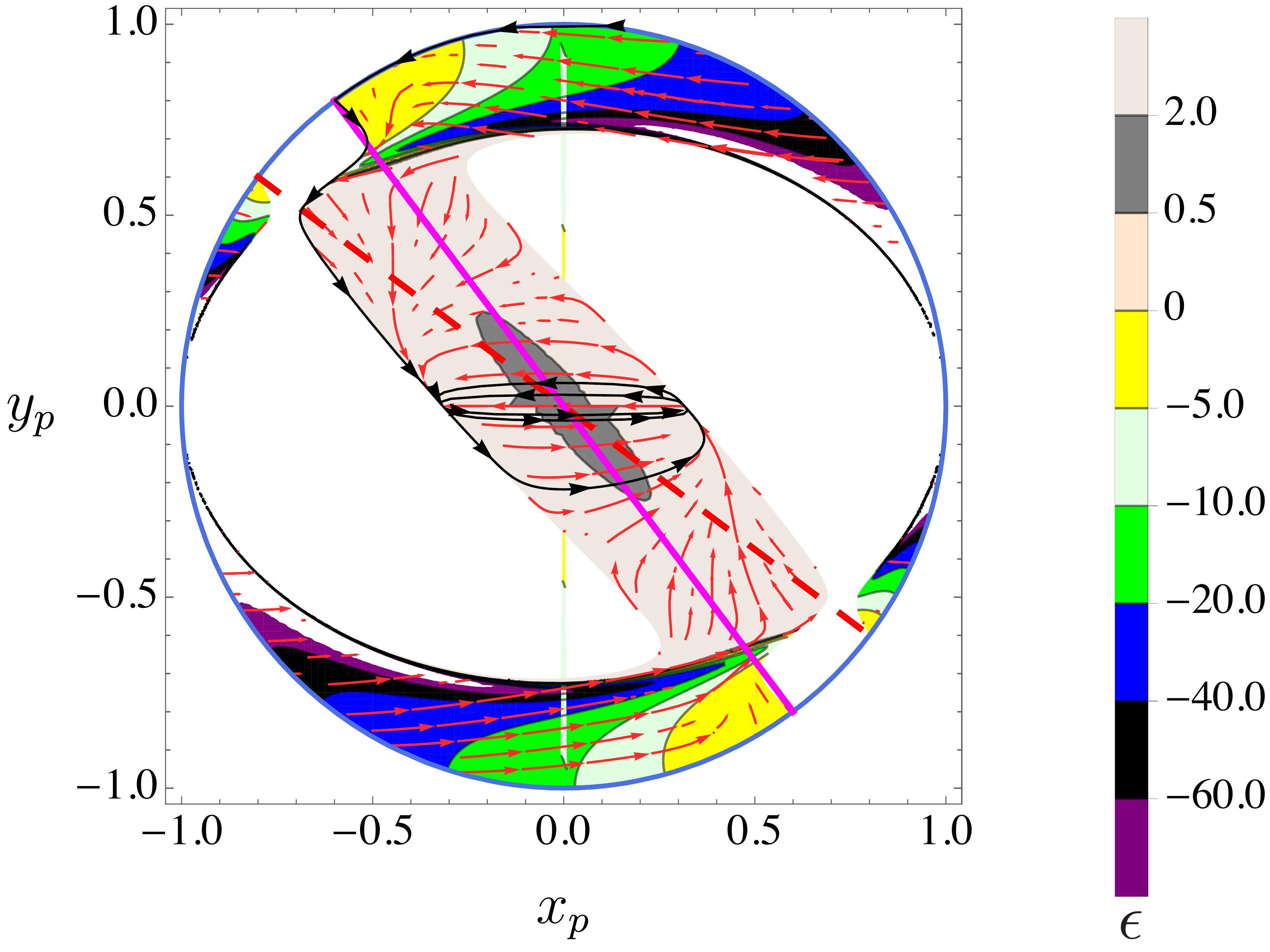} }}%
    \caption{Slow-roll inflation in the GSU2P.  This variety of primordial inflation is given when the attractor straight line of constant-roll inflation overlaps the straight line that serves as part of the border of the allowed region where the former lies.  The parameters for this plot are $\alpha_1 = 1, \alpha_3 = 1.00005$, and $\chi_5 = -160 \alpha_1 + 158 \alpha_3$.  The numerical solution in black arrows has initial conditions $x_{\rm ini} = 10^{23}$ and $y_{\rm ini} = 10^{24}$.  The zoom parameter is $\zeta = 25$ in the left plot and $\zeta = 3$ in the right one.  Contours of $\epsilon$ are drawn and distinguished by a colour code.}%
    \label{fig8}%
\end{figure}

For single-field slow-roll inflation in Einstein gravity, the second slow-roll condition, $|\eta| \ll 1$, is intimately related to the condition
\begin{equation}
|\ddot{\phi}| \ll |3 H \dot{\phi}| \,,
\end{equation}
which makes the Klein-Gordon equation behave as
\begin{equation}
3 H \dot{\phi} \approx - V'(\phi) \,,
\end{equation}
where the prime in the above expression means a derivative with respect to the scalar field $\phi$ and $V$ is the scalar potential.  In the context of the scenario discussed in this paper, one might think naively that, according to the Eq. (\ref{croll}), slow-roll inflation is met when $\beta \rightarrow -\infty$.  This is not the case, however, since the analogous to the Klein-Gordon equation gets modified in the presence of modified gravity, see Eq. (\ref{psifield}).  The first slow-roll condition, $|\epsilon| \ll 1$, is clearly satisfied in the constant-roll scenario of this paper because the inflation is de Sitter.  Along the straight line in phase space corresponding to the constant-roll inflation, the $\eta$ parameter acquires the value
\begin{equation}
\eta = \frac{3 (160 \alpha_1 - 158 \alpha_3 + \chi_5)}{2 (80 \alpha_1 - 86 \alpha_3 - 3 \chi_5)} \,,
\end{equation}
which clearly becomes zero when the equality condition in Eq. (\ref{windowok}) is satisfied.  Fig. \ref{fig6} shows contours of $\eta$ in all the plots except in the upper left where $\eta = 0$ along the curve.  Thus, the slow-roll variety of inflation is given when the attractor straight line that represents constant-roll inflation matches that one that serves as part of the border of the allowed region which the former resides in.  Fig. \ref{fig8} presents the phase space in the Poincar\'e circle for primordial slow-roll inflation together with a particular numerical solution in black arrows.  $\alpha_1 = 1, \alpha_3 = 1.00005$, and $\chi_5 = -160 \alpha_1 + 158 \alpha_3$ in the plots of this figure with the zoom parameter $\zeta = 25$ in the left plot and $\zeta = 3$ in the right one.  As can be seen in the figure, this period of primordial inflation is followed by a radiation dominated stage.

\subsubsection{Ultra slow-roll inflation}

Eqs. (\ref{fried}), (\ref{accel}), and (\ref{psifield}) reveal that the modified Klein-Gordon equation in the GSU2P is given by:
\begin{equation}
f(\psi) \ddot{\psi} + g(\psi,\dot{\psi}) 3 H \dot{\psi} + h(\psi) = 0 \,,  \label{ultra}
\end{equation}
where $f,g$ and $h$ are specific functions of their respective parameters that are not shown here.  $h(\psi)$ reduces to the derivative of a potential function with respect to $\psi$ in the limit of Einstein gravity and, therefore, is the relevant function in ultra slow-roll inflation.  In this variety of primordial inflation, the potential gets extremely flat, too flat for the slow-roll conditions to apply, i.e. $h(\psi) = 0$.  Ultra slow-roll was introduced in Refs. \cite{Kinney:2005vj,Martin:2012pe,Namjoo:2012aa,Mooij:2015yka,Romano:2016gop} and later shown to be very useful to generate the conditions for the production of primordial black holes \cite{Germani:2017bcs}.  In terms of the dimensionless variables $x,y$, and $p$, Eq. (\ref{ultra}) looks like
\begin{equation}
\tilde{f}(y) p + \tilde{g}(y,x) x + \tilde{h}(y) = 0 \,,
\end{equation}
where $\tilde{f},\tilde{g}$ and $\tilde{h}$ are minor modifications to the functions $f,g$ and $h$ respectively.  This expression, after implementing the constant-roll condition in Eq. (\ref{croll}), looks in turn like
\begin{equation}
x \left[\frac{\sqrt{2} \tilde{f}(y)}{\beta} + \tilde{g}(y,x) \right] + \tilde{h}(y) = 0 \,.
\end{equation}
Thus, in the GSU2P, the condition $h(\psi) = 0$, i.e., $\tilde{h}(y) = 0$, is formally obtained when $x = 0$ or $\sqrt{2}\tilde{f}(y)/\beta + \tilde{g}(y,x) = 0$.  In the former case, the slope of the constant-roll straight line is negative infinite which implies each of the terms in Eq. (\ref{ultra}) becomes zero;  this situation corresponds to the lower bound for $\chi_5$ in Eq. (\ref{windowok}).  The latter case, for a high value of ${\rm max}(|x|,|y|)$, can only be obtained if 
\begin{equation}
-80\alpha_1 + 86\alpha_3 + 3\chi_5 = 0 \,,
\end{equation} 
which is inconsistent with the parameter window described in Eq. (\ref{windowok}).  The only possibility is, thereby, $x = 0$, i.e. $\beta = -\infty$.  Ref. \cite{Dimopoulos:2017ged} demystified ultra slow-roll inflation showing that the name is not actually representative of what happens in GR since the scalar field in this situation flies over the flat patch of the potential.  In the case of the GSU2P, however, ultra slow-roll inflation is so ultra slow-roll that the field $\psi$ does not move at all ($\dot{\psi} \propto x = 0$).  It is worthwhile emphasizing that this is the situation when the condition $h(\psi) = 0$ is formally satisfied.  If the latter condition is relaxed and, therefore, a very large (in absolute value) but negative finite slope is chosen, the field does move slowly ($\dot{\psi} \propto y/\beta = x \approx 0$), again in stark contrast with the GR situation.  This recalls what happens in scalar-tensor theories where the ultra slow-roll inflation is actually ultra slow \cite{Motohashi:2019tyj}.  Thus, in the GSU2P, the conditions seem to be more similar to those of standard ultra slow-roll inflation in scalar-tensor theories the larger the absolute value of the slope is.  However, as is shown in the following, the closest scenario to ultra slow-roll inflation is actually not ultra slow roll.

\begin{figure}%
    \centering
    \subfloat[\centering The closest scenario to ultra slow-roll inflation \label{fig9a}]{{\includegraphics[width=9cm]{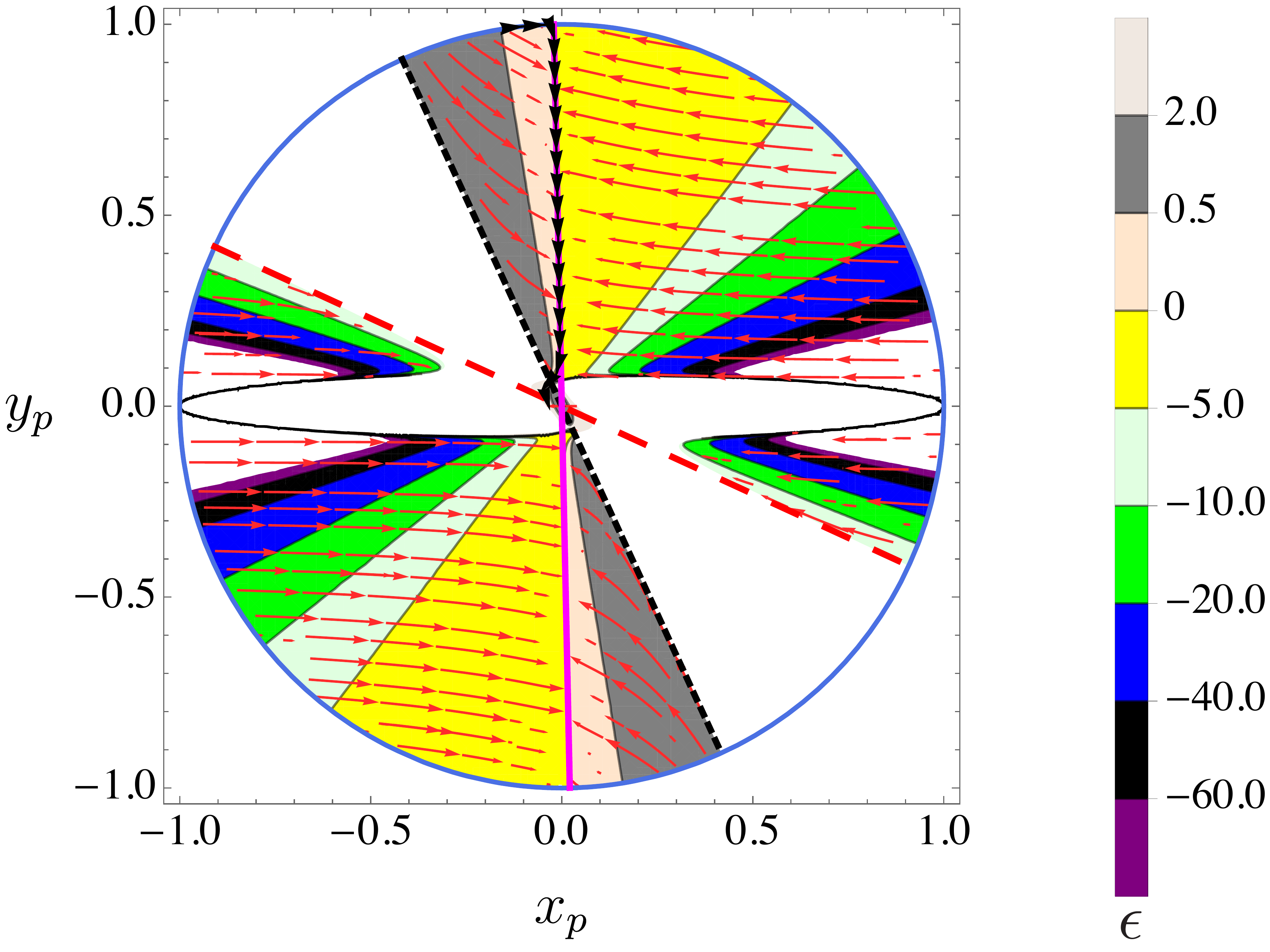} }}%
    \qquad
    \subfloat[\centering The closest scenario to ultra slow-roll inflation \label{fig9b}]{{\includegraphics[width=9cm]{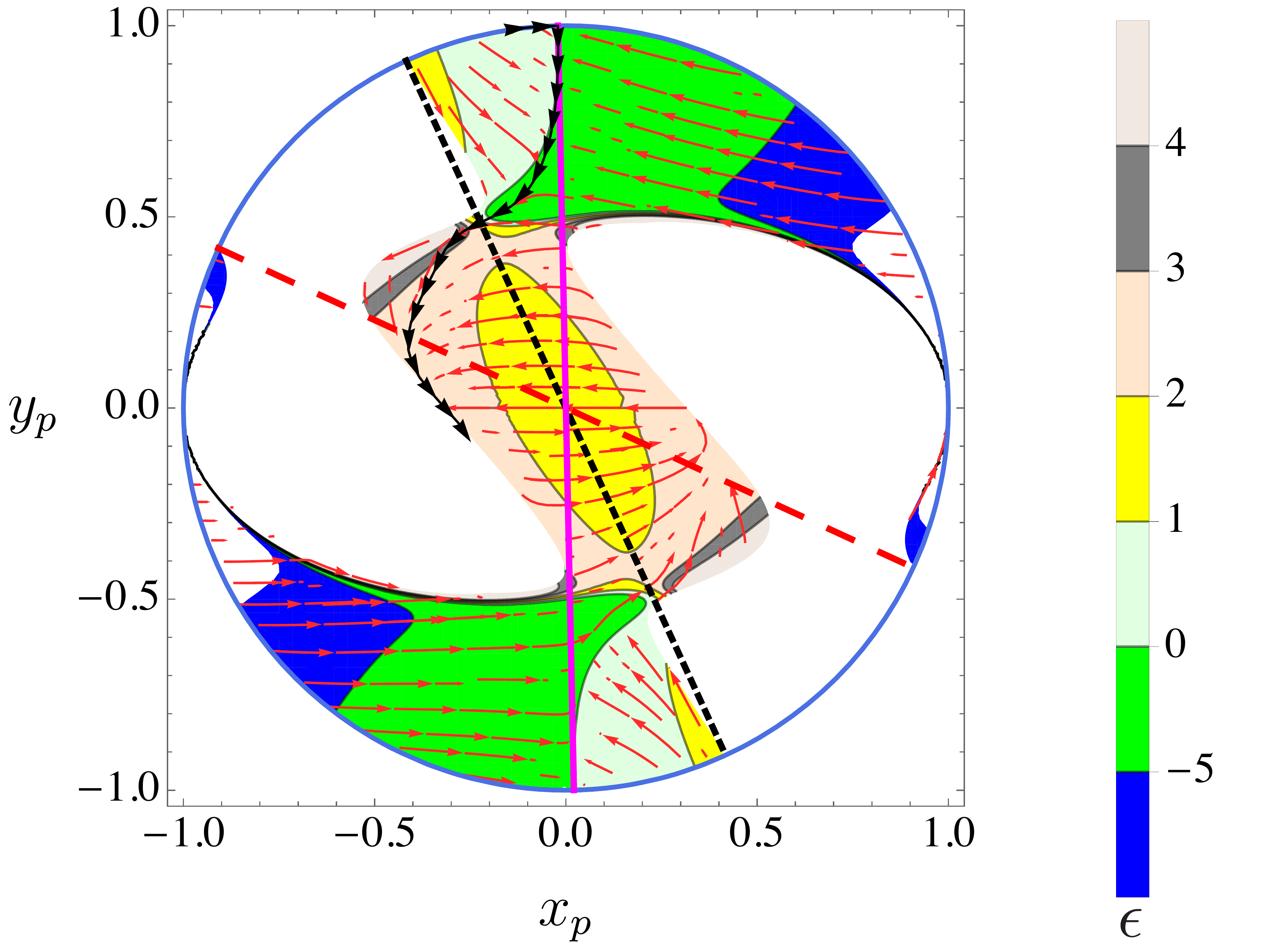} }}%
    \caption{The closest scenario to ultra slow-roll inflation in the GSU2P.  Ultra slow roll is formally given when the slope of the attractor straight line of constant-roll inflation is negative infinite. The inflation in this situation is so ultra slow-roll that the field, actually, does not evolve.  This is the reason why the plots in this figure have been drawn with a very large (in absolute value) but negative finite slope, representing the closest scenario to ultra slow-roll.  The parameters are $\alpha_1 = 1, \alpha_3 = 1.001$, and $\chi_5 = -1.981$.  The numerical solution in black arrows has initial conditions $x_{\rm ini} = -10^2$ and $y_{\rm ini} = 10^3$;  difficulties with the numerical integrator are the reasons why the damped oscillations in the central zone are not shown.  The zoom parameter is $\zeta = 25$ in the left plot and $\zeta = 3$ in the right one.  Contours of $\epsilon$ are drawn and distinguished by a colour code.}%
    \label{fig9}%
\end{figure}

\begin{figure}%
    \centering
    {\includegraphics[width=9cm]{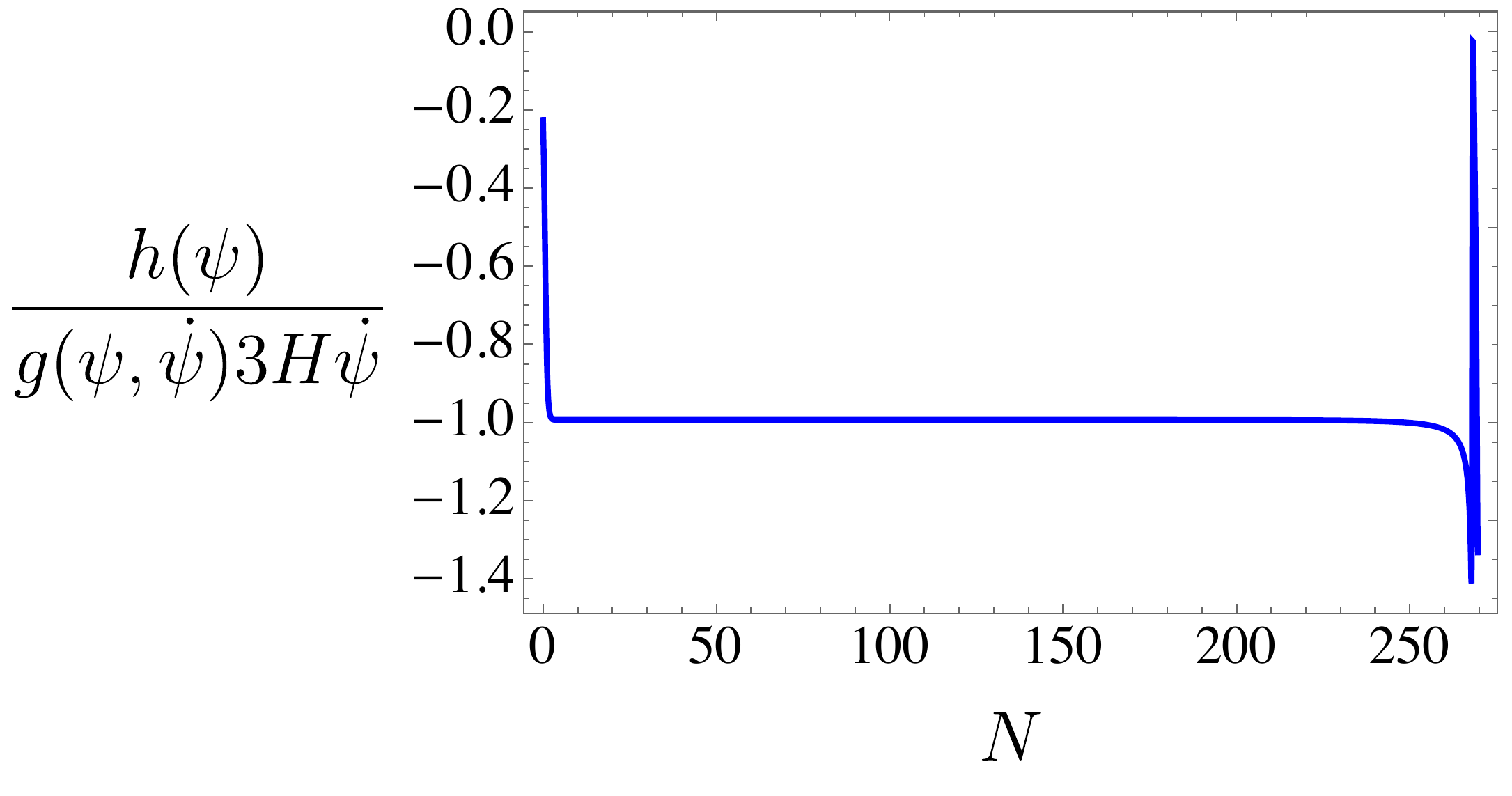} }%
    \caption{Ratio of the potential term to the friction term in the modified Klein-Gordon equation of Eq. (\ref{ultra}) for the closest scenario to ultra slow-roll inflation of Fig. \ref{fig9}.  The friction term is locked to the potential term, not to the acceleration term, which means that the closest scenario to ultra slow-roll inflation is actually not ultra slow roll.}%
    \label{fig20}%
\end{figure}

Fig. \ref{fig9}  presents two plots of the phase space in the Poincar\'e circle for the closest scenario to ultra slow-roll inflation, one with zoom $\zeta = 25$ and the other with $\zeta = 3$.  The parameters chosen for the construction of the two plots are $\alpha_1 = 1, \alpha_3 = 1.001$, and $\chi_5 = -1.981$.  A particular numerical solution is shown with initial conditions $x_{\rm ini} = -10^2$ and $y_{\rm ini} = 10^3$.  In this setup, primordial inflation is also followed by a radiation dominated period.  Ultra slow-roll inflation in the GSU2P is formally obtained when the third term in Eq. (\ref{ultra}) vanishes which, as seen before, implies the vanishing of both the first and the second terms of this equation.  Therefore, in the closest scenario to ultra slow roll, the three terms in Eq. (\ref{ultra}) are expected to be non vanishing but very small.  Moreover, the third term is expected to be much smaller than the other two.  What happens however, as seen in Fig. \ref{fig20}, is that the third term in Eq. (\ref{ultra}) is minus the second term in a very good approximation which is reminiscent of slow-roll inflation.  As a result, the closest scenario to ultra slow-roll inflation is actually not ultra slow roll but it is not slow roll either, the reason for the latter being that the absolute value of the $\eta$ parameter is not much smaller than 1 as can be observed in Fig. \ref{fig6}.

Thus, although slow roll corresponds to a limiting case in the constant-roll inflation scenario, ultra slow roll does not.  Slow roll is realized when the slope of the straight line of constant-roll inflation acquires the smallest possible absolute value before the constant-roll straight line moves outside the allowed region in phase space while ultra slow-roll  is formally realized when the slope is negative infinity.  This aspect of constant roll vs slow roll and ultra slow roll had already been stated in Ref. \cite{Motohashi:2014ppa}.  However, a negative infinity slope for the straight line of constant-roll inflation implies the field does not move at all.  Moreover, the closest scenario to the ideal ultra slow-roll inflation shows that the friction term in the modified Klein-Gordon equation is actually locked to the potential term, not to the acceleration term which would be the identifying trace of ultra slow-roll inflation.  This variety of inflation is, therefore, not present in the constant-roll inflationary scenario in the GSU2P.

\subsection{Past singularities}

Fig. \ref{fig4a} is characteristic of the evolution of the Hubble parameter in the GSU2P.  It reveals that all the integral curves at large ${\rm max}(|x|,|y|)$ come from the edge of the Poincar\'e circle so that the farther they are from the central zone, the smaller (the greater) but positive the Hubble parameter is when the flow comes from the right (the left) of the attractor straight line of constant-roll inflation.  This implies that, previous to primordial quasi de Sitter inflation, the universe experiences super accelerated expansion (either accelerated expansion or decelerated expansion that turns, later on, to accelerated expansion) and, unless $\dot{a} = 0$ at some time, and there is no reason for it to happen, there cannot exist any Big-Bang singularity.  

Another compelling singularity within the context of GR is found at the center of black holes. Interestingly, classical Y-M theories provide a natural setting for non-singular solutions at the center of black holes due to the (magnetic monopole) self-interaction part of the strength tensor \cite{Volkov:1989fi,Volkov:1998cc}.  Thus, non-Abelian vector fields seem to be equipped with an inherent mechanics to treat with singularities.

\section{Comparison with other scenarios discussed in the literature}  \label{comparison}

Several works on the role of vector fields and/or scalar fields during the primordial inflationary period or the late one can be found in the literature.  Some of the most interesting and more related to the work presented in this paper are briefly summarized in the following and the differences are highlighted.

\begin{enumerate}

\item {\it The Sushkov's paper, Ref. \cite{Sushkov:2009hk}}:

Sushkov works with the term $G_{\mu \nu} \nabla^\mu \phi \nabla^\nu \phi$ plus kinetic term plus Einstein-Hilbert in the context of the Horndeski theory.  His results are similar to those presented in this paper but, in contrast, they depend on the sign of the coupling constant.  The most probable reason for this is that $G_{\mu \nu} \nabla^\mu \phi \nabla^\nu \phi$ and the kinetic term are comparable for high values of the scalar field $\phi$.  For small $\phi$, the kinetic term becomes the leading one and, as expected, the system ends up describing a stiff fluid.

\item {\it G-inflation, Refs. \cite{Kobayashi:2011nu,Kobayashi:2010cm,Kamada:2010qe}}:

G-inflation is also framed in the context of the Horndeski theory (or Galileon theory) and shows a similar behaviour to the work presented in this paper.  Although the exit of inflation is not quite evident from the action, a stiff fluid stage is reached as happens in Sushkov's work.  Reheating is given through the gravitational production of particles.  The generalized version of the scenario \cite{Kobayashi:2011nu} shows that $\dot{\phi}$ is constant in contrast to the work presented in this paper.

\item {\it GP applied to cosmology, Refs. \cite{DeFelice:2016yws,DeFelice:2016uil,Nakamura:2017dnf}}:

There is a huge difference between the GP and the GSU2P applied to cosmology because the starring role in the GP is played by the non-propagating degree of freedom whereas in the GSU2P, as already discussed in the introduction, the propagating degrees of freedom are the protagonists.  In the Horndeski theory, the protagonist degree of freedom does propagate;  this could explain why the results in the GSU2P are similar to those found in the Horndeski theory.  Moreover, the de Sitter epoch in GP is an attractor so it only works to describe dark energy, and $\psi$ (the norm of the physical vector field) grows while $H$ decreases;  all of this is in strong contrast to what happens in the Horndeski theory and the GSU2P.

\item {\it The Emami, Mukohyama, Namba, and Zhang's paper, Ref. \cite{Emami:2016ldl}}:

Emami et. al.'s paper considers terms belonging to $\mathcal{L}_2$ and $\mathcal{L}_4$ in a version of the GP consisting of three copies of the theory where the vector fields are artificially oriented so that they form a cosmic triad and the action is invariant under global SO(3) (for a version of this scenario with non-vanishing intrinsic curvature for the spatial slices, see Ref. \cite{Murata:2021vnb}).  There exist several differences with the scenario presented in this paper.  In contrast to Emami et. al.'s paper, none of the terms of the form $B_\mu B_\nu A^{\mu \rho} A^\nu_{\;\; \rho}$ is absolutely necessary in the GSU2P to guarantee the absence of ghosts (actually for the system to behave as GR does up to second order in the tensor perturbations); however, a term of the form $B^2 A^2$ is necessary in the GSU2P (at the end, one of the terms of the form $B_\mu B_\nu A^{\mu \rho} A^\nu_{\;\; \rho}$ is required in the GSU2P but for another reason:  to sustain inflation).  Emami et. al.'s scenario requires $\chi_3 \neq 0$ (translating from their notation to this paper's) but this is not possible if the gravity waves propagate at the right speed;  anyway, Emami et. al.'s paper appeared before the constraints from GW170817 \cite{TheLIGOScientific:2017qsa,GBM:2017lvd,LIGOScientific:2017zic} did.  It is possible that the ghosts could have been avoided in Emami et. al.'s paper without having to introduce $\mathcal{L}_2^3$ if they had introduced beyond Proca terms.   It is worth mentioning that Emami et. al.'s paper considers a situation where $\dot{\psi} = 0$, $\psi$ being the norm of the cosmic triad, in order to obtain a quasi de Sitter stage which is most probably related to a critical point in phase space instead of to an attractor curve;  indeed, the quasi de Sitter stage is only possible if $\dot{\psi} = 0$, which is significantly different to the dynamics presented in this paper and possibly has to do with the absence of self interaction in Emami et. al.'s work.  

\item {\it The Adak, Akarsu, Dereli, and Sert's paper, Ref. \cite{Adak:2016led}}:

Adak et. al. obtain an inflationary scenario from a variant of $\mathcal{L}_6$, with constant coupling function, in the GP.  They look for a setup different to the usual one where the spatial components of the vector field are turned off which, of course, gives way to anisotropies.  In contrast to the work presented in this paper, the scenario of Adak et. al.'s does not have a graceful exit of inflation and, so, its inflationary solution looks more like an attractor.  Their scenario is quite similar to Einstein-Aether \cite{Jacobson:2004ts} in the sense that an extra constraint allows the coupling $Y(R) F^2$, where $Y(R)$ is an arbitrary function of the Ricci scalar and $F$ is the Faraday tensor, without generating higher-order derivatives.  In this sense, inflation is given by a power law that is near to de Sitter.  Overall, the proposal of Adak et. al.'s exhibits some similarities to the work presented in this paper.

\item {\it HYM-flation, Refs. \cite{Oliveros:2019zkl,Davydov:2015epx}}:

Oliveros et. al.'s  \cite{Oliveros:2019zkl} and Davydov et. al.'s \cite{Davydov:2015epx} scenarios, both of them being variants of what could be dubbed Horndeski Y-M inflation (HYM-flation), look like extensions to $\mathcal{L}_6$ of Emami et. al.'s scenario.  Whereas Oliveros et. al. implement the isotropy by making the action invariant under local SO(3), Davydov et. al. do it by making the action invariant under local SU(2).  This makes the two scenarios essentially the same although the amount of inflation is different; the reason for this seems to be the self interaction in Davydov et. al.'s paper that is absent in Oliveros et. al.'s.  Although searching for quasi de Sitter inflation, Oliveros et. al.'s and Davydov et. al.'s papers consider $\dot{\psi} = 0$ as Emami et. al. do; this is, again, significantly different to the dynamics presented in this paper.

\item {\it The Shahidi's paper, Ref. \cite{Shahidi:2018sas}}:

Shahidi's paper is quite particular because it involves a scalar and a vector field but, because of the form of the action, the vector field contributes only at the perturbative level.  As in other proposals described above, de Sitter inflation in Shahidi's scenario is obtained for a constant value of the scalar field $\phi$, i.e., Shahidi's paper is dealing with a critical point.

\item {\it The Hrycyna's paper, Ref. \cite{Hrycyna:2020jmw}}:

The system under study in Hrycyna's work is constituted of a scalar field $\phi$ conformally coupled to gravity in the form $\phi^2 R$.  Hrycyna's paper presents an idea that is quite similar to the one discussed in this paper because he analyzes critical points assuming very large values for $\phi$.  This is done for a given region where $\phi$ is larger than some critical value and, therefore, all the critical points can be constituted as asymptotic states.  In analogous fashion to the work presented in this paper, Hrycyna's paper shows that $\dot{\phi} \rightarrow - \infty$ when $\phi \rightarrow \infty$ in the de Sitter asymptotic state.

\end{enumerate}

\section{Conclusions} \label{conclusions}

Primordial inflation in the GSU2P displays nice features that have deserved detailed analysis.  Constant-roll de Sitter inflation stands out as an attractor curve in phase space whose attraction basin covers almost all the allowed region, making inflation in this scenario a really predictive phenomenon.  The slow-roll variety of inflation presents itself as a limiting case of the constant-roll dynamics.  The inflation may be made, in all the cases, long enough to solve the classical problems of the standard cosmology, it being followed by a graceful exit into a radiation dominated period.  The interplay between the canonical kinetic (Y-M) term for the vector field and the other terms in the GSU2P, specifically those that go as a fourth power of the vector field, is crucial for the dynamics of this scenario.  This is because the latter, when they dominate, are the responsible for the primordial inflation whereas the former gives way to the radiation dominated stage when the norm of the cosmic triad has become small enough.  In addition, the scenario can be made free of ghosts and Laplacian instabilities, there does not exist any Big-Bang singularity, and the gravity theory reduces to GR when the vector field decays.  Finally, although the norm of the cosmic triad can take super Planckian values, the energy scale during inflation may be as small as wished since it is controlled by the generalized SU(2) group coupling constant.

The scenario presented in this paper has a non-void available parameter window as can be seen in Fig. \ref{fig7}.  Notwithstanding the latter being quite small, it is clear that the GSU2P has a great potential to be home of the constant-roll de Sitter primordial inflationary period:  fourteen free parameters were too much to do a reasonable discussion about the properties of the inflationary scenario in this paper, but with all of them available, the possibilities of having a much wider available parameter window are certainly high.  Moreover, because of the technical difficulties at handling hundreds of Lagrangian building blocks, the GSU2P was partially constructed in both Refs. \cite{Allys:2016kbq} and \cite{GallegoCadavid:2020dho} considering up to six space-time indices in the Lagrangian building blocks previous to contractions with the metric tensor and/or the orientability 4-form.  Thus, should the theory be completed some day, it will promote the fourteen coupling constants to coupling functions of the vector field, and include some new additional terms as well, which means even more possibilities to enlarge the available parameter window.  Therefore, what has been presented in this paper is a proof of concept of what the GSU2P can make to describe inflation and, of course, of the great potential it has to describe other cosmological phenomena. 

The background physics has been the most important subject of this paper but, clearly, in order to put the inflationary scenario under test, the full perturbative treatment is required.  The forthcoming work will consist of calculating the primordial curvature perturbation, its spectrum and bispectrum, and comparing the predictions of this scenario on the tensor-to-scalar ratio - spectral index plane vs the Planck data \cite{Planck:2018jri} and the level of non-gaussianity with the upper bound given by Planck \cite{Planck:2019kim}.  Because of the symmetries of the problem at hand, it is possible to anticipate that the perturbations of the three members of the cosmic triad are equal to each other in average;  this means that not only the expansion is isotropic but the different spectra are statistically isotropic \cite{Rodriguez:2015xra}.  The scenario discussed in this paper can be reasonably thought, therefore, as an effective single-scalar-field mechanism of primordial inflation.  The vector signatures of the scenario will reveal when the cosmic triad be slightly modified so that the anisotropies in the expansion \cite{Maleknejad:2011jr,Orjuela-Quintana:2020klr}, in the spectrum \cite{Dimopoulos:2008yv}, and in the bispectrum \cite{Karciauskas:2008bc,Almeida:2014ava} of the primordial curvature perturbation can be calculated.  These aspects as well as the potential of this scenario to explain the formation of primordial black holes, as described in Ref. \cite{Motohashi:2019rhu}, will also be investigated.

\medskip
\medskip

\medskip
\textbf{Acknowledgments} \par 
Y. R. benefitted from some discussions and correspondence with Diana L\'opez Nacir, Filippo Vernizzi, and Paolo Creminelli.  L. G. G acknowledges financial support from Vicerrector\'{\i}a de Investigaci\'on, Desarrollo e Innovaci\'on - Universidad de Santiago de Chile, project DICYT,
Grant No. 042031CM$\_$POSTDOC.  Y. R. has received funding/support from the Patrimonio Aut\'onomo - Fondo Nacional de Financiamiento para la Ciencia, la Tecnolog\'{\i}a y la Innovaci\'on Francisco Jos\'e de Caldas (MINCIENCIAS - COLOMBIA) Grant No. 110685269447 RC-80740-465-2020, project 69553 as well as from the European Union's Horizon 2020 research and innovation programme under the Marie Sklodowska-Curie grant agreement No 860881-HIDDeN.  Maple, Mathematica, and the package xAct (\url{http://www.xact.es}) were employed to perform and cross-check all the analytical and numerical calculations carried out in this paper.

\medskip

%
\bibliographystyle{MSP}
\bibliography{Bibli.bib}





\end{document}